\documentclass[aps,pre,amsmath,twocolumn,amssymb,nofootinbib]{revtex4}
\usepackage[colorlinks,citecolor=blue,bookmarks]{hyperref}
\usepackage{graphicx,amssymb,rotate}
\usepackage{mathrsfs}
\usepackage{color}
\usepackage{multirow}
\usepackage{float}
\usepackage{pstricks}
\usepackage[toc,page]{appendix}
\usepackage[mathscr]{euscript}
\usepackage{textcomp}
\usepackage{hhline}
\def\ADNDT{{At. Data Nucl. Data Tables }}

\def\NP{Nucl. Phys. A }
\def\PR{Phys. Rev. }
\def\PRL{Phys. Rev. Lett. }

\def\PRC{Phys. Rev. C }
\def\NDT{{Nucl. Data Tables }}
\usepackage{rotating}
\usepackage[labelformat=parens,labelsep=quad, skip=3pt]{caption}
\usepackage{subfig}

\def\etal{\!{\em et al.\ }}

\def\arrow{$\rightarrow$}

\begin{document}
\title{
                  Bound and continuum state $\beta^-$ decay of bare atoms: enhancement of decay rate and changes in $\beta^{-}$ decay branching}

\author{Arkabrata Gupta}  
\author{Chirashree Lahiri}
\author{S. Sarkar}
 \thanks{ Corresponding author: ss@physics.iiests.ac.in}
\affiliation {Department of Physics, Indian Institute of Engineering Science and Technology (Formerly, Bengal Engineering and Science University), Shibpur, Howrah-711103, India}

\today
\begin{abstract}

We have calculated rates of $\beta^{-}$ decay to both continuum and bound states separately for some fully ionized (bare) atoms in the mass range A $\approx$ 60-240. One of the motivations of this work is that the previous theoretical calculations were very old and/or informatically incomplete. Probably no theoretical study on this subject has been done in the last three decades. For the calculation, we have derived a framework from the usual $\beta^{-}$ decay theory used by previous authors. Dependence of the calculated rates on the nuclear radius and neutral atom Q-value have been examined. We have used the latest experimental data for nuclear and atomic observables, such as $\beta^{-}$ decay Q-value, ionization energy, neutral atom $\beta^{-}$ decay branchings, neutral atom half-lives etc. Results of $\beta^{-}$ decay rates for decay to continuum and bound states and the enhancement factor due to the bound state decay for a number of nuclei have been tabulated and compared with the previously calculated values, if available. The effective rate or half-life calculated for bare atom might be helpful to set a limit for the maximum enhancement due to bound state decay. Finally, $\beta^{-}$ decay branching for bare atom has been calculated. The changes in branching in bare atom compared to that in the neutral atom and for the first time branching flip for a few cases have been obtained.
Reason for this branching change has been understood in terms of Q-values of the transitions in the neutral and bare atoms. Verification of this branching change / flip phenomenon in bare atom decay might be of interest for future experiments. 

\end{abstract}

\maketitle

\section{Introduction}

It is well known that the usual theory of $\beta^{-}$ decay presumes that the decay of a neutron to proton is accompanied by the creation of an electron and an anti-neutrino in continuum states. However, in a stellar plasma where atoms get partially or fully ionized, this continuum decay is not the sole option. Nuclear $\beta ^{-}$ decay to the bound states of the ionized atom is another probable channel. Also bare atoms have been produced terrestrially and $\beta ^{-}$ decays have been studied in storage ring experiments. In 1947 Daudel \etal \cite{daudel} first proposed the concept of bound state $\beta$ decay. This suggests that a nucleus has a possibility to undergo $\beta ^{-}$ decay by creating an electron in a previously unoccupied atomic orbital instead of the continuum decay. It is important to understand that the bound state decay process does not occur subsequently from the $\beta^{-}$ decay of an electron previously created in the continuum state, it is rather the direct creation of an electron in an atomic bound state accompanied by a mono-energetic anti-neutrino created in the free state carrying away the total decay energy. This process has been studied both theoretically as well as experimentally over the past seven decades. 

In case of a neutral atom, available phase space for the creation of an electron in a vacant atomic orbital is very small and therefore the bound state decay is almost negligible compared to the contribution of the continuum decay. Contrarily, ionization of atoms may lead to drastic enhancement of bound state $\beta$ decay probability due to the availability of more unoccupied atomic levels. In some previous theoretical works from 60's to 80's, various groups have studied the continuum and bound state  $\beta$ decay for neutron, tritium \cite{bahcall} and  fully ionized (bare) heavy atoms  \cite{takahashi, takahashi1, takahashi2}. However, in most cases, previous theoretical works were based on very old data and/or informatically incomplete. Simultaneously, the development of experimental techniques has served fruitfully to detect bound and continuum state $\beta$ decay channels of fully ionized atoms. In 1992, Jung \etal first observed the bound state $\beta^{-}$ decay for the bare $^{163}$Dy atom \cite{jung} by storing the fully ionized parent atom in a heavy-ion storage ring. In the same decade, Bosch \etal studied the bound state  $\beta ^{-} $ decay for fully ionized $^{187}$Re \cite{bosch} which was helpful for the calibration of $^{187}$Re - $^{187}$Os   galactic chronometer \cite{yokoi}. Further experiments with bare $^{207}$Tl \cite{ohtsubo} showed the simultaneous measurement of bound and continuum state $\beta^{-}$ decay. However, the authors have mentioned this decay as a single $\beta^{-}$ transition process to a particular daughter level with 100 \%  branching \cite{ohtsubo} whereas, the present data \cite{nndc} suggest three available levels among which the total $\beta^{-}$ decay is distributed. 
 
In earlier studies, Takahashi and Yokoi \cite{takahashi, takahashi2} had investigated $\beta$ transition (bound state $\beta^{-}$ decay and orbital electron capture) processes of some selected heavy nuclei suitable for s-process studies. However, in their work, they had not given separately the bound state decay rate of bare atoms. Further, in another work, Takahashi  \etal \cite{takahashi1} had studied the $\beta^{-}$ decay of some bare atoms for which bound state $\beta^{-}$ decays produce significant enhancement in decay rates and proposed measurement in storage ring experiment. 
However, they did not take into account  the contribution of transitions to all possible energy levels of the daughter nucleus in total $\beta^{-}$ decay rate enhancement. As an example, according to the present $\beta^{-}$ decay data \cite{nndc}, there are six possible  $\beta^{-}$ transitions from the [117.59 keV, $6^{+}$] state of $^{110}$Ag to various states of $^{110}$Cd, but they had mentioned the contribution of only one transition.


With the availability of modern day experimental $\beta$ decay half-lives in terrestrial condition for the neutral atom, experimental  Q-values  for $\beta ^{-} $ decays and atomic physics inputs, it becomes inevitable to re-visit some of the earlier works. Moreover, in a previous work, Takahashi and Yokoi \cite{takahashi} addressed a few  nuclei in their `case studies', undergoing $\beta ^{-}$ transitions, as some of the essential turnabouts in $s$-process nucleosynthesis, where contributions from atoms with different states of ionization were considered. However, the explicit study of bound and continuum state $\beta ^{-}$ transitions of bare atoms for most of these nuclei remained unevaluated till date both experimentally as well as theoretically.

In the present work, our aim is to study the $\beta ^{-} $ decay of some elements, in the mass range (A $\approx$ 60 - 240) which might be of interest for future experimental evaluations using storage ring. In particular, calculations of $\beta^{-}$ decay rates to the continuum as well as bound state of these fully ionized atoms, where information for neutral atom experimental half-life and  $\beta^{-}$ decay branchings are terrestrially available, have been performed. Most importantly the study of effective half-lives for bare atoms will be helpful to set a limit for the maximum enhancement in $\beta^{-}$ decay rate due to the effect of bound state decay channels.  
Moreover, we have also discussed the effect of different nuclear structure and decay inputs (Q value, radius etc.) over the bound to continuum decay rate ratio. In addition, some interesting phenomena of changes in $\beta^{-}$ decay branching for a number of bare atoms along with some notable change in branching (branching-flip) for a few of them, have been obtained. The branching-flip is obtained for the first time.


The paper is organized as follows: section \ref{2} contains the methodology of our entire calculation for bound and continuum state $\beta^{-}$ decay rates for bare atom,  as well as comparative half-life ($Log ft$) for neutral atom. In section \ref{3A} we have discussed our results for the neutral atoms, whereas in section \ref{3B}  results for the bare atoms have been discussed. The phenomenon of change in $\beta^{-}$ decay branching for bare atom compared to that in neutral atom is also discussed explicitly in the section \ref{3B}.  Conclusion of our work has been described in section \ref{4}. Finally, we present a table for the calculated $\beta^-$ decay rates in Appendix A followed by a discussion on the choice of spin-parity for unconfirmed states of neutral atom in Appendix B.


\section{{Methodology} \label{2}}

 In this work, we have dealt with the allowed and first-forbidden $\beta ^{-} $ transitions for neutral and fully ionized atoms. The contributions of higher-order forbidden transitions are negligible in the determination of the final $\beta ^{-} $ decay rate and thus we have not tabulated the contributions for the same.    

The transition rates (in $sec^{-1}$) for allowed (a), non-unique first-forbidden(nu) and unique first-forbidden(u) transitions are given by \cite{takahashi, takahashi1, takahashi2}

\begin{eqnarray}
\lambda = [(ln 2)/f_0t](f^{*}_{m})     ~~~~~~ \text {for m= a, nu}  \\ \nonumber
=[(ln 2)/f_1t](f^{*}_{m})~~~~~~ \text{for m= u ~~      }.
\end{eqnarray}

Here $t$ is the partial half-life of the specific parent-daughter energy level combination for which transition rate has to be calculated and $f^{*}_{m}$ is the lepton phase volume part described in detail, below in this section. For allowed and non-unique first-forbidden $\beta ^{-} $ decay, the expression for the decay rate function $f_0(Z,W_0)$ can be simplified to \cite{gove, konopinski} 

\begin{eqnarray}
f_0(Z,W_0) =\int^{W_0}_1\sqrt{(W^2-1)} W (W_0-W)^2\\ \nonumber
\times F_0(Z,W)L_0dW .
\end{eqnarray}

The certain combinations of electron radial wave functions evaluated at nuclear radius R ( in the unit of $\hslash/m_ec$) were first introduced by Konopinski and Uhlenbeck \cite{konopinski} as $L_k$'s. The value for $k=0$ can be approximated  as
 
\begin{equation}
L_0 = \dfrac{1+\sqrt{1-\alpha^2 Z^2}}{2}.
\end{equation}
Here, $\alpha$ is the fine structure constant. In the work of Behrens and J\"anecke \cite{behrens}, the authors had taken a different form of $L_0$, which includes a slight dependence on the momentum. However, we find that the $L_0$ approximation, adopted in our calculation, is in  good agreement with that from the Ref. \cite{behrens} within the considered energy window.  

In Eq.(2), $W$ is the total energy of the $\beta^{-}$ particle for a $Z-1\rightarrow Z $ transition and $W_0 = Q_n/m_ec^2+1 $ is the maximum energy available for the $\beta^{-}$ particle. Here the mass difference between initial (parent) and final (daughter) states of neutral atoms are expressed as the decay $Q$ value ($Q_n$ in keV). The term $F_0(Z,W)$ is the Fermi function for allowed and non-unique transition, given by \cite{konopinski}

\begin{eqnarray}
F_0(Z,W)=\dfrac{4}{ \left|\Gamma \left( {1+2\sqrt{1-\alpha^2 Z^2 }}\right)\right|^2}\\ \nonumber
\left(2R\sqrt{W^2-1}\right)^{2\left(\sqrt{1-\alpha^2 Z^2} -1\right)}exp\left[\dfrac{\pi \alpha Z W}{\sqrt{W^2-1}}\right] \\ \nonumber
\times \left|{\Gamma{\left(\sqrt{1-\alpha^2 Z^2} + i \dfrac{\alpha Z W}{\sqrt{W^{2}-1}} \right)}}\right |^2.
\end{eqnarray}

Similarly, for the unique first-forbidden transition the decay rate function $f_1(Z,W_0)$ has the form reduced from Refs. \cite{gove, konopinski} is given by,

\begin{eqnarray}
f_1(Z,W_0) = \int^{W_0}_1\sqrt{(W^2-1)} W (W_0-W)^2 F_0(Z,W) \\ \nonumber
\times\left[(W_0-W)^2  L_0 + 9 L_1 \right]dW ,
\end{eqnarray}

with $L_1$ given by,

\begin{equation}
L_1 = \dfrac{F_1(Z,W)}{F_0(Z,W)} \left(\dfrac{W^2-1}{9}\right) \dfrac{2+\sqrt{4-\alpha^2 Z^2}}{4}.
\end{equation}
 
The term $F_1(Z,W)$ for unique $\beta ^{-} $ transition is given by \cite{konopinski},

\begin{eqnarray}
F_1(Z,W)=\dfrac{(4!)^2}{ \left|\Gamma \left( {1+2\sqrt{4-\alpha^2 Z^2 }} \right) \right|^2} \\ \nonumber
\left(2R\sqrt{W^2-1}\right)^{2\left(\sqrt{4-\alpha^2 Z^2} -2\right)}exp\left[\dfrac{\pi \alpha Z W}{\sqrt{W^2-1}}\right] \\ \nonumber
\times \left|{\Gamma{\left(\sqrt{4-\alpha^2 Z^2} + i \dfrac{\alpha Z W}{\sqrt{W^{2}-1}} \right)}}\right |^2.
\end{eqnarray}

Eqs. (2) and (5) are general forms of $f_0(Z,W_0)$ and $f_1(Z,W_0)$. For more precise calculation of f-factor, one should in principle, include various corrections in the integrand of Eqs. (2) and (5). Corrections due to atomic physics effects, radiative correction and finite nuclear size effects might be important for such studies.  For fully ionized atoms, corrections due to atomic physics effects, such as, imperfect overlap of initial and final atomic wave functions, exchange effects that comes from the anti-symmetrisation of the emitted electron with the atomic electrons \cite{bahcall2}, screening of the nuclear charge due to the coulomb field of the atomic electronic cloud are not needed. For neutral atom, the decay to the atomic bound state should be negligible \cite{bahcall2}. Also, the screening and exchange corrections together cancel a large part of the overlap correction \cite{budick}. Further the non-orthogonality effect becomes rapidly smaller as Z increases \cite{takahashi1}. Some of the corrections have positive sign and some of them have negative sign. So unless all the corrections are taken together, the treatment for corrections to f- factor will not be consistent. Therefore we have neglected these contributions both for bare and neutral atoms. We have included the correction due to the extended charge distribution of the nucleus on the $\beta^{-}$ spectrum. This correction is $\Lambda_k(Z,W) \rightarrow \Lambda_k(1+\Delta\Lambda_k(Z,W)$), where the term $\Lambda_k$ can be written in terms of $L_k$ and $F_0(Z,W)$ as \cite{gove, konopinski}

\begin{equation}
\Lambda_k(Z,W)= F_0(Z,W)L_{k-1}\left[ \dfrac {(2k-1)!!}{(\sqrt{W^2-1})^{k-1}}\right]^2 ,
\end{equation}
  
in such a way that it reduces to $\left[ F_0(Z,W)L_0\right]$ and $\left[ 9F_0(Z,W)L_1/(W^2-1)\right]$ for $k=1$ and $2$, respectively. The correction term is given by \cite{gove},

\begin{eqnarray}
\Delta\Lambda_k (Z,W) =(Z-50) \times \\ \nonumber
\left[ -25\times 10^{-4} - 4\times10^{-6} W \times (Z-50)\right] \\ \nonumber
\text {   for  } k = 1, Z > 50 , \\ \nonumber
= 0 ~~~~~~~~~~~~~~~~~~~~~    \text {   for   } k = 1 , Z \le 50, \\ \nonumber
= 0 ~~~~~~~~~~~~~~~~~~~~~~    \text {for                      } k > 1~~~~~~~~~~~.
\end{eqnarray}

The screened energy of the emitted electron $(W')$ enters through $\Delta\Lambda_k(Z,W')$, where $W'=W-V_0(Z)$. We calculated $V_0(Z)$, following Gove and Martin \cite{gove}, using expression from W. R. Garrett and C. P. Bhalla \cite{bhalla}. This correction to the integrand in Eqs. (2) and (5) has effect in the fourth decimal place of the f-factor and this is consistent with Ref. \cite{hayen} discussed for the allowed $\beta^{-}$ decay. So we have dropped $W'$ and used $W$ in the integrand.

It is to be noted that in the present work we have used experimental quantities, such as Q - value, half-life, branching, which have uncertainties even up to the first decimal place \cite{nndc, nist}. So, in our treatment we have neglected the screening effect too for neutral atom. Therefore, by using Eqs. (8) and (9)  in the integrand of Eq. (2) and Eq. (5) one can calculate the values for  $f_0(Z,W_0)$ and $f_1(Z,W_0)$  incorporating  only finite size correction.

 In the work of  A. Hayes \etal \cite{hayes}, the authors have taken a different form of the finite-size correction  involving the charge density, which has a complicated radial dependency. However, we find that the results from the present calculation are in agreement with the available experimental data \cite{nndc}. 

 Further, from the above expressions (Eqs.(4) and (7)), it is evident that the factors $F_0(Z,W)$ and $F_1(Z,W)$  depend on the radius, thereby making the terms $f_0$ and $f_1$ (Eqs.(2) and (5)), radius dependent. Thus, in our present study, we have used various radius values from different phenomenological models and experiments to study their effects on the final $ft$ values. In order to calculate $ft$ values for a nucleus, we have extracted the half-life $t$ for individual transition to daughter levels using the latest $\beta$ decay
branching information available in the literature \cite{nndc}.

The lepton phase volume $f^{*}_{m}$ \cite{takahashi2} for the continuum state  $\beta^{-}$ decay can thus be expressed as,

\begin{eqnarray}
f^{*}_{m=a,nu}(Continuum) = \int^{W_c}_1\sqrt{(W^2-1)} \\ \nonumber
 W (W_c-W)^2 F_0(Z,W) L_0  dW,
\end{eqnarray}
 
and
 
\begin{eqnarray}
f^{*}_{m=u}(Continuum) = \int^{W_c}_1\sqrt{(W^2-1)} \\ \nonumber
W (W_c-W)^2 F_0(Z,W) \times \\ \nonumber
 \left[(W_c-W)^2 L_0 + 9L_1\right]  dW,
\end{eqnarray}

Here $W_c = Q_c/m_ec^2 +1 $ is the maximum energy available to the emitted $\beta^{-}$ particle, and $Q_c$ is given by,

\begin{eqnarray}
Q_c = Q_n - \left[ B_n(Z+1) - B_n(Z)\right].
\end{eqnarray}
     
 The term $\left[ B_n(Z+1) - B_n(Z)\right]$ denotes the difference of binding energies for bound electrons of the daughter and the parent atom. The experimental values for all the atomic data (binding energies/ionization potential) are availed from Ref. \cite{nist}. 

Further, for the bound state  $\beta^{-}$ decay of the bare atom  $f^{*}_{m}$ takes the form \cite{takahashi2}

\begin{eqnarray}
f^{*}_{m=a,nu}(Bound) = \sum_x \sigma_x \left(\pi/2\right) \left[ f_x\text{ or }g_x \right]^2 b^2 \\ \nonumber
\left(\text {for  } x=ns_{1/2},np_{1/2}\right),
\end{eqnarray}
 
and
 
\begin{eqnarray}
f^{*}_{m=u}(Bound) = \sum_x \sigma_x \left(\pi/2\right) \left[ f_x\text{ or }g_x \right]^2 b^4  \\ \nonumber
\left(\text {for  } x=ns_{1/2},np_{1/2}\right), \\ \nonumber
~~~~~~~~~~~~~~~~~~~~\\ \nonumber
= \sum_x \sigma_x \left(\pi/2\right) \left[ f_x \text{ or } g_x \right]^2 b^2 \left(9/R^2\right) \\ \nonumber
\left(\text{for  } x=np_{3/2},nd_{3/2}\right).
\end{eqnarray}

Here $\left[ f_x\text{ or }g_x \right]$ is the larger component of electron radial wave function evaluated at the nuclear radius $R$ of the daughter for the orbit $x$. The $\left[ f_x\text{ or }g_x \right]$ is obtained by solving Dirac radial wave equations using the Fortran subroutine RADIAL by Salvat \etal \cite{cpc}. In our case,  $\sigma_x$ is the vacancy of the orbit, chosen as unity and $b$ is equal to $Q_b/m_ec^2$ where, 

\begin{eqnarray}
Q_b = Q_n - \left[ B_n(Z+1) - B_n(Z)\right]-B_{shell}(Z+1).
\end{eqnarray}
 
For example, in case of a bare atom, if the emitted $\beta^{-}$ particle gets absorbed in the atomic K  shell, then the last  term of Eq.(15) will be the ionization potential for the K electron denoted by $B_K(Z+1)$. 


\section{Results and Discussion}

In this work, we have calculated  
$\beta^-$ decay transition rates to  bound and continuum states,  for a number of  fully ionized atoms in the mass range A $\approx$ 60-240. One of the motivations is that there are some evidences where earlier works were not equipped enough to address the entire $\beta^-$ decay scenario.
  This might be due to the unavailability of information about all the energy levels participating in transition processes.

As an example, Takahashi \etal \cite{takahashi1} have considered transitions for allowed(a), first-forbidden non-unique(nu) and first-forbidden unique(u) decay of parent nuclei to a few energy levels of daughter nuclei. For instance, in the case of $^{228}$Ra nucleus, the authors have tabulated the decay from the ground state of the parent $[E$(keV), $J^\pi] = [0.0, 0^+]$ nucleus to $[6.3, 1^-]$ and $[33.1, 1^+]$ states of the daughter nucleus  $^{228}$Ac. However, these two transitions cover only the 40\% of the total $\beta^{-}$ decay branching of neutral $^{228}$Ra atom from the ground state. With the latest experimental data \cite{nndc}, we find that there are two more available states of  $^{228}$Ac where the rest amount of $\beta^{-}$ decay from the ground state of $^{228}$Ra occur. In this section, it will be shown that the contributions of all these four states are extremely important in the determination of effective enhancement of $\beta^{-}$ transition rates of bare $^{228}$Ra as well as to understand the phenomenon of  branching-flip, discussed in section \ref{3B}.

 For simplicity, this section is subdivided into two parts. The first subsection involves the calculation of  $Log~ ft$ for the neutral atom, a necessary ingredient for the calculation of $\beta^-$ decay rate of the bare atom. In the next subsection, the $\beta^{-}$ decay transition rates of  bare atoms have been discussed with a detailed explanation of TABLE {\red A.I}. The dependence of these decay rates on different parameters is also  examined in the same subsection. Finally, we have shown and discussed the change in individual level branchings in fully ionized atoms.

\subsection{{ $Log~ ft$ calculation for neutral atoms}\label{3A}}

It is evident from Eqs.(1-9) that the calculation for $ft=f_0t (/f_1t)$ is one of the essential components in the determination of the transition rate $\lambda$, which in turn depends on radius R of the daughter nucleus. However, $Log~ ft$ data obtained from Ref. \cite{nndc} can not provide the information of the $R$ dependence of $Log~ ft$.
As the present theoretical modelling  for bare atom depends on radius (see section \ref{2}), we find it more accurate to calculate  $Log~ ft$ for neutral atom  for different choices of radii. 

In Appendix \ref{lognu}, we present a table for bound and continuum state $\beta^-$ decay rates for bare atoms along with the values of $Log~ ft$ for corresponding neutral atoms at different radii and compare our calculations with existing theoretical as well as experimental results (see the supplemental material \cite{supl} for details). 
As explained in section \ref{2}, we have tabulated $Log~ ft$ values only for allowed (a), first-forbidden non-unique (nu) and first-forbidden unique (u) transitions.

Here, in TABLE {\red A.I}, $R_1$ is the phenomenological radius evaluated as $R_1=1.2 A^{1/3}$ fm, whereas $R_2$ is the nuclear charge radius in fm \cite{angeli} and $R_3$ is the half-density radius given by \cite{gove} $R_3=(1.123A^{1/3} - 0.941A^{-1/3})$ fm. We have calculated  $Log~ ft$ values for $R_1$, $R_2$ and $R_3$ and compared them with the existing data \cite{nndc}. Besides, we have tabulated the available values from previous calculations of Takahashi \etal \cite{takahashi1} in the same table. 

One can see that the change in radius may cause a change in the $Log ~ft$ value mostly in the second decimal place. In the next subsection, we will show the effect of these variations on the transition rates for bare atoms.

Further, from TABLE {\red A.I} and the supplemental material \cite{supl}, it can be noted that our calculation matches with the experimental  $Log~ ft$ data \cite{nndc} in most cases up to the first decimal place. The agreement of our result with experimental data \cite{nndc} confirms the applicability of the methodology adopted in the present  study.

\subsection {{ Bound and Continuum decay rates of bare atoms}\label{3B}}

 In  the ninth and the eleventh column of TABLE  {\red A.I} of Appendix \ref{lognu}, bound and continuum $\beta^-$ decay rates of bare atoms are presented, respectively. 

 It is observed that the dependence on radius affect the bound ($\lambda_B$) and the continuum state ($\lambda_C$)  decay rates  in first or second decimal places, and the ratio $\lambda_B/\lambda_C$ remains almost unaffected up to the first decimal place for most of the examined cases. 

Further, from TABLE  {\red A.I} (also see the supplemental material \cite{supl}), we find that the values for $\lambda_B$ and $\lambda_C$ from our calculation agree with those of the existing theoretical results \cite{takahashi1} quite well. The possible reasons for the slight mismatch between our calculation and that from Takahashi \etal \cite{takahashi1} 
are mainly due to (i) the effect of the nuclear radius, (ii) the adoption of present day Q values (for all $Q_n ,  Q_c$ and $Q_b$), (iii) availability of present day  $\beta^-$ decay branching of neutral atoms and (iv) the choice of significant digits. Despite that, the overall success of our calculation in reproducing available $\lambda_B$ and $\lambda_C$ for bare atoms once again justify the extension of the present calculation to previously unevaluated cases.  

\begin{figure*}

{\includegraphics[width=15cm,height=11cm]{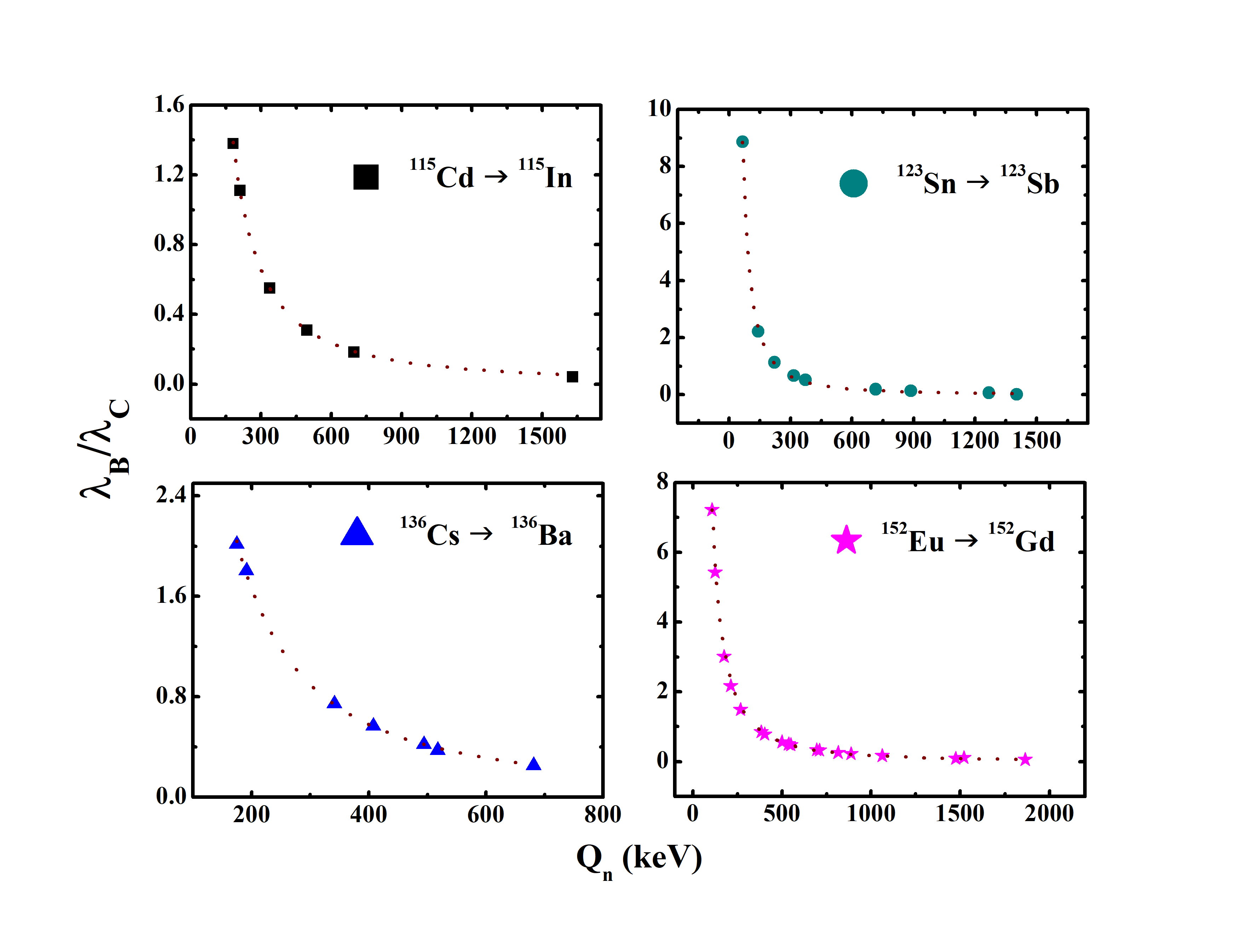}}
\caption{(Color online) Ratio of $\lambda_B/\lambda_C$ Vs the neutral atom Q-value $Q_n$ (in keV) for various $\beta^{-}$ transitions (for the radius $R_{1}$). The dotted curves are obtained from fitting to Eq.(16).  See text for details.  
\label{lblc}}
\end{figure*}

It can again be shown from TABLE  {\red A.I} that in a transition from the parent nucleus $^AX_{Z-1}$  to different energy levels of the daughter nucleus $^AX_{Z}$, the ratio $\lambda_B/\lambda_C$ for all transitions are not same, rather it decreases with increasing $Q_n$ value. It can be understood from the expressions in Eqs.(10-15) where the factors $f^{*}_{Continuum}$ and $f^{*}_{Bound}$ depend on $Q_c$ and $Q_b$, respectively, which are again derived from the neutral atom Q value $Q_n$. Due to different $Q_n$ values for different transitions,  $\lambda_B/\lambda_C$ can be identified as a function of $Q_n$. For the sake of understanding, in FIG. \ref{lblc}, we have plotted the ratio $\lambda_B/\lambda_C$ versus $ Q_n$ for the nuclei $^{115}$Cd, $^{123}$Sn, $^{136}$Cs and $^{152}$Eu. In each case, dependence on $Q_n$ is observed which can be fitted to the form
 
\begin{eqnarray}
\dfrac{\lambda_{B}}{\lambda_{C}}=a\times({Q_n})^b
\end{eqnarray}

where a and b are the nucleus dependent constants given in TABLE  \ref{ab}.

\begin{table}[H]

\caption{Parameters a and b for Eq.(16) for the radius $R_1$.}

\vspace*{0.3 cm}
	\centering
\resizebox{!}{!}{
 \begin{tabular}{|c|c|c|}
		\hline
                 Parent $\rightarrow$ Daughter        &              Parameter a           &       Parameter b \\ \hline
 \rule{0pt}{0.5 cm}
                 $^{115}Cd$ $\rightarrow$ $^{115}In$  &    3093.12 $\pm$ 317.17            & -1.48 $\pm$ 0.02   \\ \hline

 \rule{0pt}{0.5 cm}
                 $^{123}Sn$ $\rightarrow$ $^{123}Sb$  &    12657.22 $\pm$ 1515.52            & -1.73 $\pm$ 0.03   \\ \hline

 \rule{0pt}{0.5 cm}
                 $^{136}Cs$ $\rightarrow$ $^{136}Ba$  &    5178.76 $\pm$ 654.04            & -1.52 $\pm$ 0.02   \\ \hline

 \rule{0pt}{0.5 cm}
                 $^{152}Eu$ $\rightarrow$ $^{152}Gd$  &    18851.81 $\pm$ 1065.69            & -1.68 $\pm$ 0.01   \\ \hline

\end{tabular}}

\label{ab}
\end{table}


 The TABLE \ref{ab} confirms that Eq.(16) is a characteristic feature of $\lambda_{B}/ \lambda_{C}$ ratio of the bare atom with particular Z and A values. If there is a mistake in the calculation of  $f^{*}$ for $\lambda_{B}$ or $/$ and $\lambda_{C}$, then the ratio point will not fit to such a power law.

 In the fourteenth column of TABLE {\red A.I}, the ratio of $\lambda_{Bare}(=\lambda_{B} + \lambda_{C})/\lambda_{Neutral}$ (called here rate enhancement factor) has been tabulated. It is evident from these values that there must be an enhancement in the decay rate for each transitions (i.e. $\lambda_{Bare}/\lambda_{Neutral} > 1$) because of the additional bound state decay channel. 

\begin{figure}

{\includegraphics[width=85mm,height=65mm]{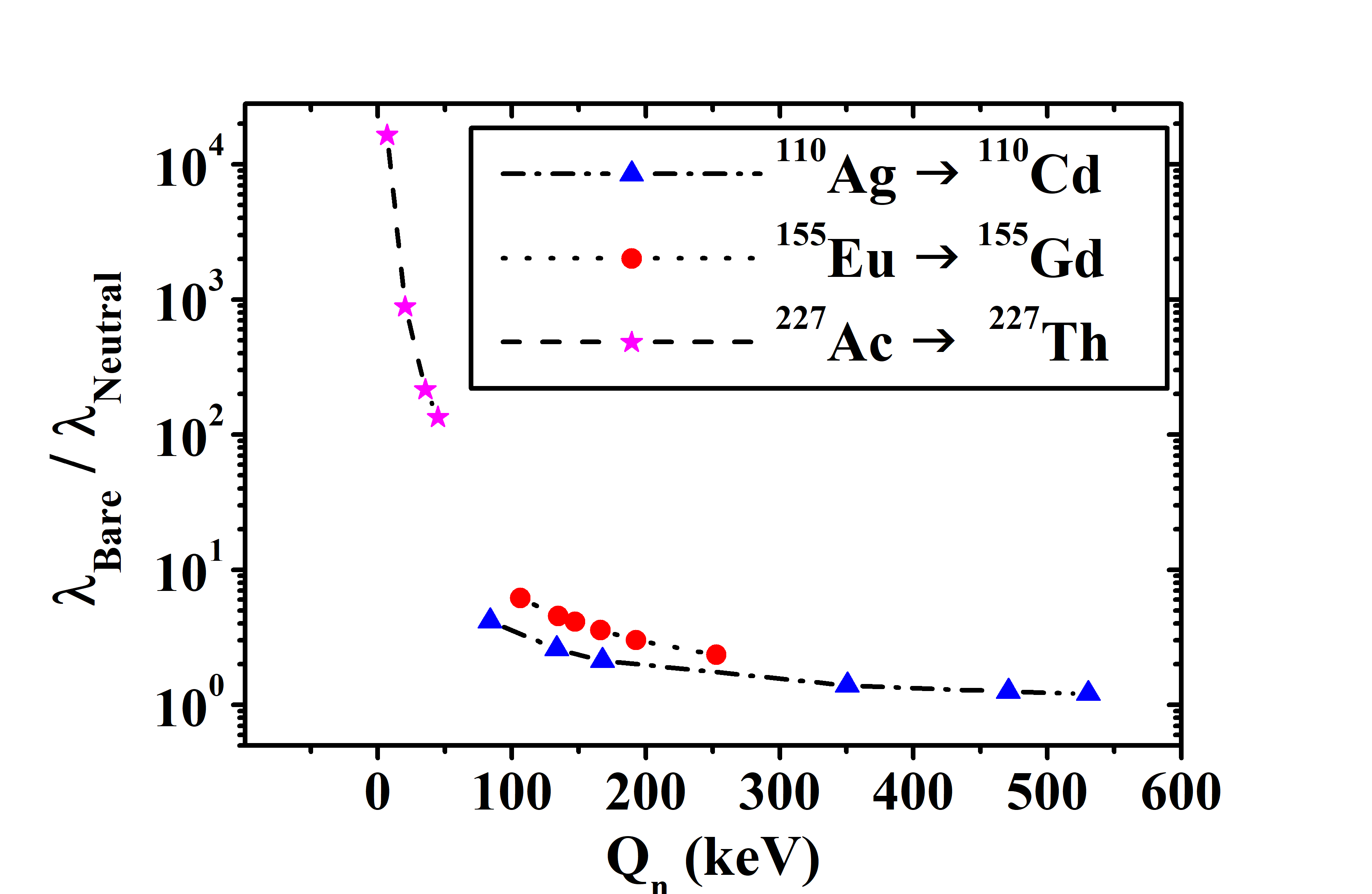}}
\caption{(Color online) Ratio of $\lambda_{Bare}/\lambda_{Neutral}$ Vs the neutral atom Q-value $Q_n$ (in keV) for various $\beta^{-}$ transitions (for the radius $R_{1}$). See text for details.  
\label{enh}}
\end{figure}

In FIG. \ref{enh}, the ratio of  $\lambda_{Bare}/\lambda_{Neutral}$  for $^{110}$Ag, $^{155}$Eu and $^{227}$Ac have been shown. From the figure, it can be noted that rate enhancements  (a) are different for different transitions of a particular nucleus, (b) are dependent on $Q_n$ values : lower the $Q_n$, larger the enhancement. Moreover, this rate enhancement factor (c) also depends on Z and A of the atom; larger the value of Z and/or A,  larger the enhancement.  

Further, in TABLE  {\red A.I}, we have tabulated effective $\beta^-$ decay half-lives for bare atoms and compared to those of neutral atoms. It should be noted that the neutral atom half-life given in the fifteenth column of the table is the total half-life corresponding to a, nu and u types of $\beta^{-}$ transitions only.

\vspace{0.2cm}

\underline {\textbf{Transition details: case studies}}  
\vspace{0.2cm}

\begin{figure*}[t]
\centering
{\includegraphics[width=19.5cm,height=7.5cm]{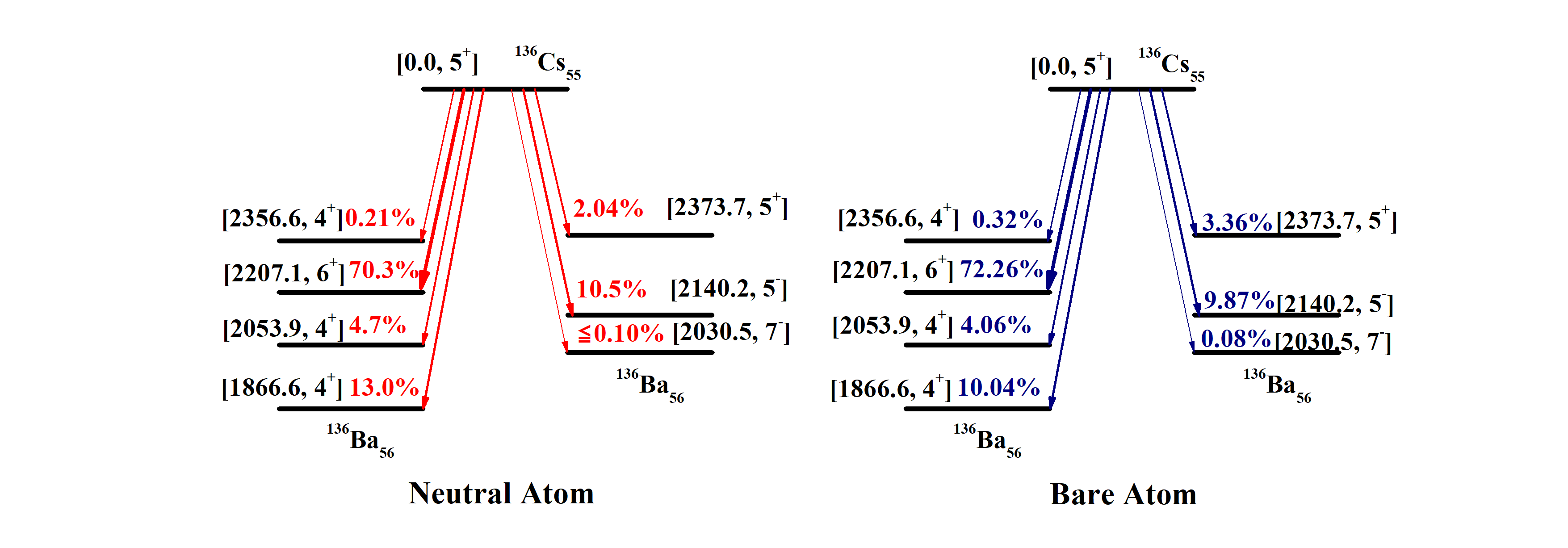}}
\caption{(Color online) Comparison of the $\beta^{-}$ decay branchings for neutral and bare $^{136}$Cs isotope (for the radius $R_{1}$). 
\label{noflip}}
\end{figure*}

The dependence of the rate enhancement factor on $Q_n$ causes a change in $\beta^-$ branching for the bare atom. In bare atom, branchings similar to the neutral atom can only be achieved if the factor $\lambda_{Bare}/\lambda_{Neutral}$ remains constant with  $Q_n$, which is obviously not the case (FIG. \ref{enh}). In other words, this change can be understood to be an outcome of the non-uniformity of the  $\lambda_B / \lambda_C $ ratio with $Q_n$. It is observed that the continuum decay rate for bare atom decreases with respect to that for the neutral atom (i.e. $\lambda_{C} < \lambda_{Neutral}$) due to the reduction of continuum Q value ($Q_c < Q_n$, Eq. (12)).  Further from FIG. \ref{lblc}, it is clear that with the decrease in the $Q_n$ value, $\lambda_B$ dominates more over $\lambda_C$ and hence the effective decay rate of the bare atom $\lambda_{Bare}=\lambda_B + \lambda_C$ does not follow the same branching as that of the neutral atom.

 Note: For the $\beta^{-}$ transition having very low $Q_n$ value, bound state decay may be the only path of $\beta^{-}$ decay. As an example, in the transition of $^{227}$Ac $[0.0,3/2^-]$ to  $^{227}$Th $[37.9,3/2^-]$ with $Q_n = 6.9$ keV, $Q_c$ for continuum decay of the bare atom becomes $-13.1$ keV. As evident from Eqs.(10-12), due to the negative value of  $Q_c$, the corresponding decay channel gets closed. On the other hand, as $(Q_b-Q_n) > 0$ for this transition, the total decay is governed by the bound state channel only.

As an example, in FIG. \ref{noflip}, we have compared branchings for  neutral (left panel) and bare (right panel) $^{136}$Cs atom. It can be seen from FIG. \ref{noflip} that the branchings for all $\beta^-$ transitions of the bare atom have been changed from that of the neutral atom. However, the ordering of each branch remains unaltered in both cases, i.e. the $[0.0, 5^+] \rightarrow [2207.1, 6^+]$ branch gets the maximum feeding followed by the $[0.0, 5^+] \rightarrow [1866.6, 4^+]$ and $[0.0, 5^+] \rightarrow [2140.2, 5^-]$ branches, whereas the minimum feed goes to  $[0.0, 5^+] \rightarrow [2030.5, 7^-]$ channel for both the neutral and bare atoms.

Further, some notable observations and comments for some nuclei are given below.  

 $\bullet$ In case of neutral $^{207}$Tl atom in terrestrial condition, the $[0.0, 1/2^+]$ state of  $^{207}$Tl  decays to $[0.0, 1/2^-]$ state of  $^{207}$Pb with 99.729\% branching, whereas to $[569.6, 5/2^-]$ state of the daughter has the branching  $>$0.00004\% (in some places of Ref. \cite{nndc} this value is given as  $<$0.00008\%) and to $[897.8, 3/2^-]$  state has 0.271\% branching \cite{nndc} (see supplemental material \cite{supl} for details). For bare atom, Ohtsubo \etal \cite{ohtsubo} had observed bound state decay rate $\lambda_B= 4.29(29) \times 10^{-4}$ sec$^{-1}$ and continuum state decay rate $\lambda_C=2.29 (012) \times 10^{-3}$ sec$^{-1}$, by considering the transition to  $[0.0, 1/2^-]$ state of $^{207}$Pb with 100\% branching. In our calculation for bare atom, we have got bound state decay rate $\lambda_B = 4.15 \times 10^{-4}$ sec$^{-1}$ and continuum state decay rate $\lambda_C= 2.54 \times 10^{-3}$ sec$^{-1}$. The calculated branchings of bare $^{207}$Tl : $\sim$ 99.6 \% to $[0.0, 1/2^-]$, $\sim$ 0.00005\% - 0.0001\% to $[569.6, 5/2^-]$ and $\sim$ 0.4 \% to $[897.8, 3/2^-]$ states of the daughter $^{207}$Pb.

\begin{figure*}[t]
\centering
{\includegraphics[width=17cm,height=6cm]{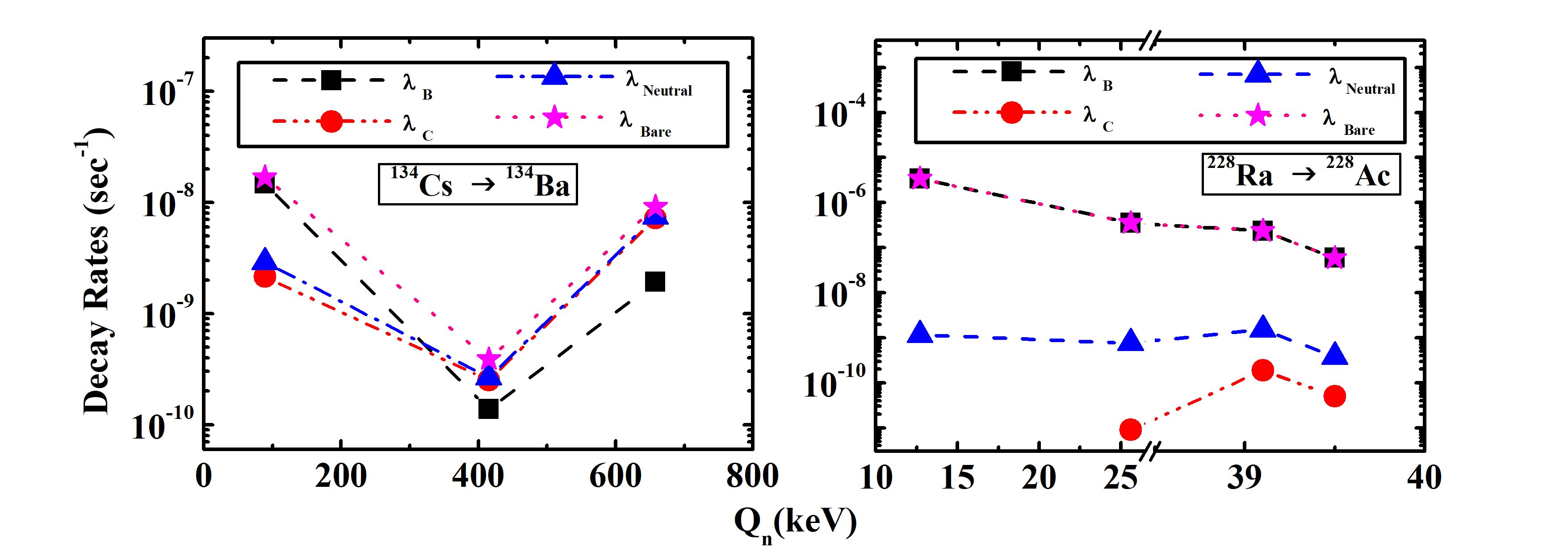}}

\caption{(Color online) Decay rates (in sec$^{-1}$) for neutral ($\lambda_{Neutral}$) and bare($\lambda_{Bare}$) atoms along with all the decay components ($\lambda_B$ and $\lambda_C$) of the bare atom (for the radius $R_{1}$) with the neutral atom Q-value $Q_n$ (in keV). See text for details.  
\label{all}}
\end{figure*}

 In our study, we found some special cases where  effective branchings for the bare atom do not follow the same ordering as that of the neutral atom. This indicates a very interesting phenomenon of branching-flip, obtained for the first time in this work. Sometimes the additive contribution of $\lambda_B$ and $\lambda_C$ and the effect of these two competing channels can lead to this branching-flip. This  can be understood from FIG. \ref{all}. In FIG. \ref{all}, decay rates (sec$^{-1}$) for neutral ($\lambda_{Neutral}$) and bare ($\lambda_{Bare}$) atom along with all decay components ($\lambda_B$ and $\lambda_C$) of the bare atom versus $Q_n$ are shown for the ground state decay of $^{134}$Cs and $^{228}$Ra nuclei. One can see from FIG. \ref{all} that the highest point corresponding to $\lambda_{Neutral}$ (i.e. maximum $\beta^{-}$ branching in neutral atom) and the highest point corresponding to $\lambda_{Bare}$ (i.e. maximum $\beta^{-}$ branching in bare atom) are coming from different transitions to the daughter nuclei (different $Q_n$ values), which clearly indicates the phenomenon of flip in the branching sequence.

$\bullet$ In the case of $^{134}$Cs,  $\lambda_{Neutral}$ is maximum at $Q_{n} = 658.1$ keV, which is due to the maximum branching to the 1400.6 keV level (see supplemental material \cite{supl} for details) of  $^{134}$Ba \cite{nndc}. In contrary, for the same nucleus, $\lambda_{Bare}$ is maximum at $Q_n = 88.8$ keV which therefore indicates the maximum branching to the 1969.9 keV level (see TABLE {\red A.I}) of the daughter $^{134}$Ba for bare atom. 

\begin{figure*}[t]
\centering
{\includegraphics[width=19cm,height=7cm]{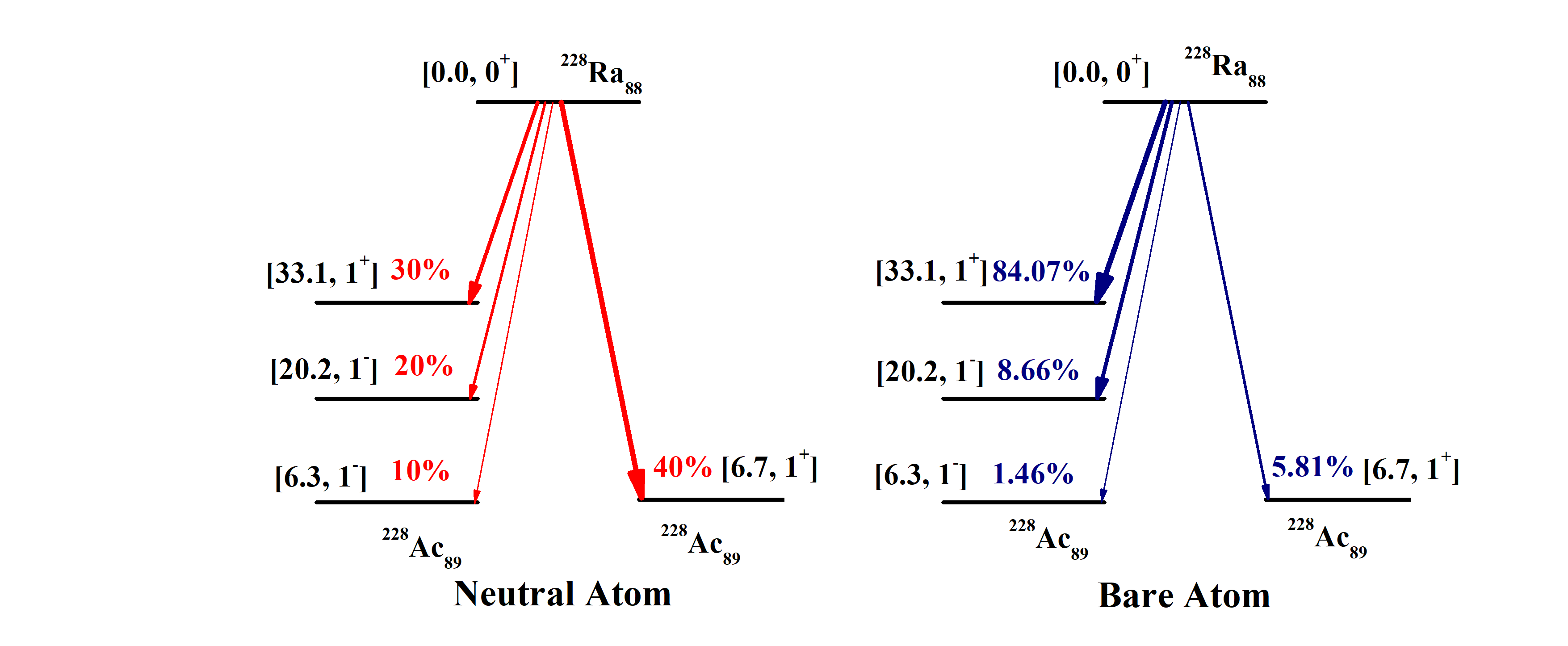}}
\caption{(Color online) Comparison of level branchings on neutral and bare $^{228}$Ra isotope (for the radius $R_{1}$). Left Panel: neutral atom, Right Panel: bare atom.   
\label{flip}}
\end{figure*}

$\bullet$ Similarly for $^{228}$Ra, the maximum branching for the bare atom ($(\lambda_{Bare})_{max}$ at $Q_n=12.7$ keV) shifts from that of the neutral atom ($(\lambda_n)_{max}$ at $Q_n=39.1$ keV). In FIG. \ref{flip}, we have shown the change and alteration of transition branchings for the $\beta^-$ decay of $^{228}$Ra.  
One can see the branching-flips of the participating levels of the $^{228}$Ac atom in FIGS. \ref{all} and \ref{flip}. In case of the neutral $^{228}$Ra atom, maximum branching is 40\% for the $[6.7, 1^+]$ level of the daughter \cite{nndc}. After complete ionization, the major contribution of the total decay rate comes due to the bound state enhancement of $Q_n=$12.7 keV channel which has $\sim$ 84.07\% decay to the  $[33.1, 1^+]$  level (30\% in neutral atom) of the daughter atom, whereas only $\sim$ 5.81\% of the total decay branching is observed for the level $[6.7, 1^+]$.   

There are a few more cases, where the branching-flips are observed. However, not necessarily, all the transition branches face the phenomenon of flip. It may also happen that only two or three branches change their sequence, whereas other branches remain in the same order as that of the neutral atom.

$\bullet$ In the $\beta^-$ decay of $^{152}$Eu $[45.5998, 0^{-}]$ (see Table 1 of the Ref. \cite{supl} for branching details), we find that in both cases (neutral and bare) the branching to  $[0.0, 0^+]$ branch of the daughter dominate over the rest, whereas a branching-flip is observed between $[344.3,2^+]$ and $[1314.6, 1^-]$ states.  

$\bullet$ Similarly for $^{227}$Ac, we find that there is a branching-flip between two transitions from   $[0.0, 3/2^-]$ state of the parent to  $[0.0, 1/2^+]$ and $[24.5, 3/2^+]$ states of the daughter atom.
The ratio of branching for these two levels is 5.4:1 for neutral atom, which changes to 1:1.38 for bare atom.

It should be noted that the ultimate fate of individual branchings in the bare atom is decided by two factors: the initial branching (required to calculate $Log~t$ for each transition: a part of $Log~ft$ calculation) and Q value of the neutral atom. The competition between these two factors determines whether the branching-flip will occur or not.

\textbf{Effect of uncertainties: } Furthermore, in order to get the complete picture of  $\beta^-$ decay for bare atom, effects due to uncertainties in $\beta^-$ decay half-life  and Q value need to be considered. The effect of uncertainty is appreciable depending on the numerical value of the half-life and Q value. In case of atoms with the  $\beta^-$ decay half-life of the order of seconds/minutes and having high Q value, no significant change is observed in the calculation of $Log~ ft$ due to the inclusion of experimental uncertainties. The contributions peek out for long lived nuclei  with large uncertainty or for transitions of high Q value having large uncertainty.   
 For example, in case of $^{93}$Zr atom, where the neutral atom half-life is equal to $1.61 \times  10^6(5)$ years, $Log~ ft$ for the transition $[0.0,5/2^+ \rightarrow 30.8,1/2^-]$ with the radius $R_1$ is given by ${10.234}^{+0.014}_{-0.013}$. Therefore, the final values for continuum and bare state $\beta^-$ transitions including the uncertainties can be written as  $\lambda_C={6.87}_{-0.21}^{+0.22} \times 10^{-15}$ $sec^{-1}$ and $\lambda_B={6.13}_{-0.19}^{+0.20} \times 10^{-15}$ $sec^{-1}$, respectively. 

\section{{Conclusion}\label{4}}
To summarize, in this work we have calculated individual contributions of bound and continuum state $\beta^-$ decays to the effective $\beta^-$ decay rate of the bare atom in the  A $\approx$ 60 to 240 mass range where earlier information were partial and/or old.

 Additionally, the dependence of transition rates over the nuclear radius and the Q value is illustrated clearly in the present study.  We found a power law dependence of $\lambda_{B}/ \lambda_{C}$ of a bare atom on $Q_n$ for each value of Z and A. Along with the effective enhancement of transition rates, we have found that transition branchings for the bare atom differs from that of the neutral atom for all Z and A, which is an outcome of non-uniform enhancement amongst the participating branches. Most interestingly, we have found few nuclei, viz. $^{134}$Cs, $^{228}$Ra etc., where some flip in the branching pattern is found for their bare configuration. It will be interesting to see how these results help planning new experiments involving bare atoms.  The calculations will be extended to partially ionized atoms which will provide decay rate as function of density and temperature of the stellar plasma and will be useful for calculation of nucleosynthesis processes.


\section*{ACKNOWLEDGEMENT}
AG is grateful to DST-INSPIRE Fellowship (IF160297) for providing financial support. CL  acknowledges the grant from DST-NPDF (No. PDF/2016/001348) Fellowship.








\appendix 

\setcounter{equation}{0}  
\section{Table for $\beta^-$ decay}\label{lognu}
\begin{figure*}
\subfloat[]{\label{Figure6:a}\includegraphics[width=6.5cm,height=6.6cm]{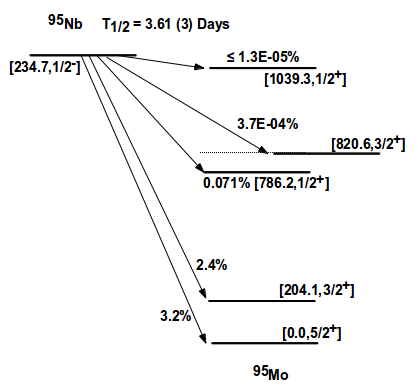}}\label{a1}\hspace*{3.00cm}
\subfloat[]{\label{Figure6:b}\includegraphics[width=6.7cm,height=6.3cm]{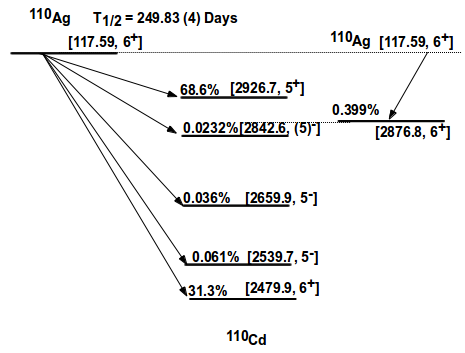}}\label{a2}\\
\vspace{.5cm}
\subfloat[]{\label{Figure6:c}\includegraphics[width=6.9cm,height=7.0cm]{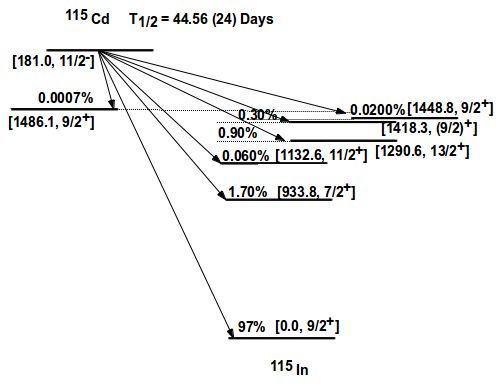}}\label{a3}\hspace*{3.00cm}
\subfloat[]{\label{Figure6:d}\includegraphics[width=6.5cm,height=6.6cm]{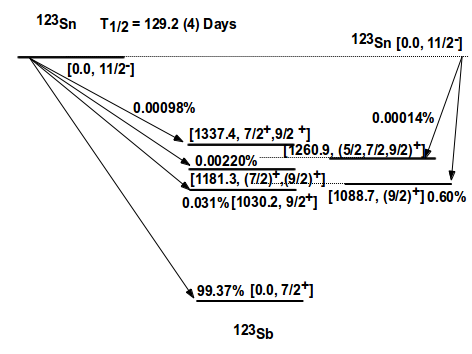}}\label{a4}\\
\vspace{1.cm}
\subfloat[]{\label{Figure6:e}\includegraphics[width=5.6cm,height=5.4cm]{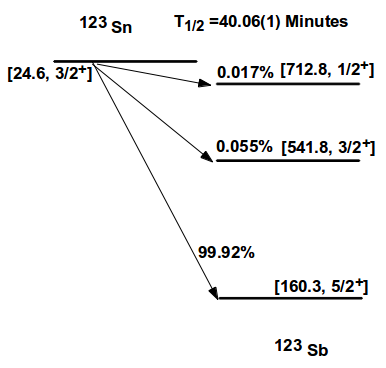}}\label{a5}\hspace*{3.00cm}
\subfloat[]{\label{Figure6:f}\includegraphics[width=5.4cm,height=3.8cm]{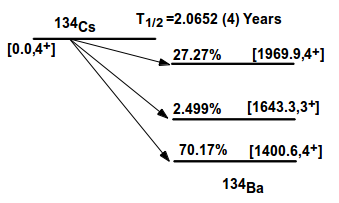}}\label{a6}\\
\caption{$\beta^-$ decay transition ($^{95}Nb$, $^{110}Ag$, $^{115}Cd$, $^{123}Sn$, $^{134}Cs$) with neutral atom branchings  are given \cite{nndc}. Here, T$_{1/2}$  is  the total half-life of the  parent  level (including all possible decay channels, viz. $\beta$, $\alpha$, IT etc).  However, only allowed (a), first-forbidden non-unique (nu) and first-forbidden unique (u) $\beta^-$ decay transitions are shown in these figures.}
\label{Figure6}
\end{figure*}
\begin{figure*}
\subfloat[]{\label{Figure7:a}\includegraphics[width=6.4cm,height=5.6cm]{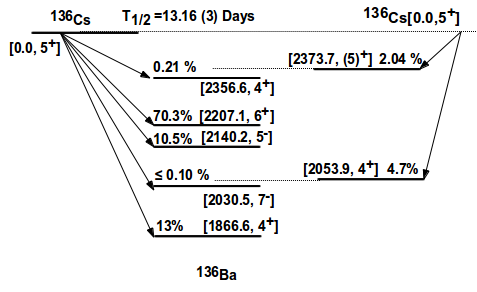}}\label{a7}\hspace*{3.00cm}
\subfloat[]{\label{Figure7:b}\includegraphics[width=7.2cm,height=7.0cm]{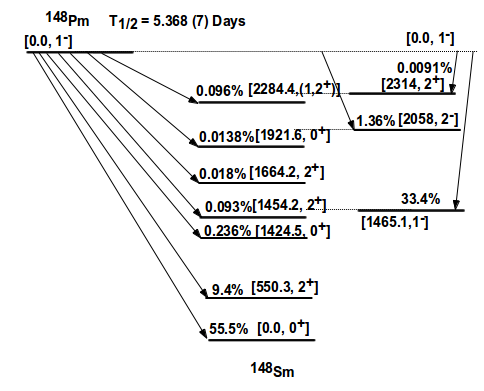}}\label{a8}\\
\subfloat[]{\label{Figure7:c}\includegraphics[width=6.6cm,height=6.0cm]{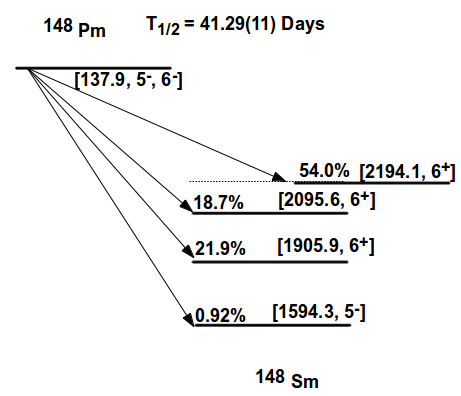}}\hspace*{3.00cm}
\subfloat[]{\label{Figure7:d}\includegraphics[width=7.0cm,height=7.2cm]{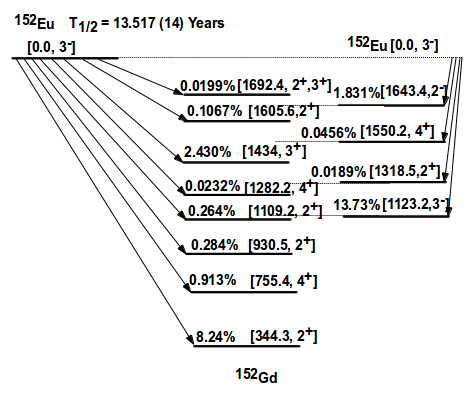}}\label{b2}\\
\subfloat[]{\label{Figure7:e}\includegraphics[width=7.3cm,height=6.8cm]{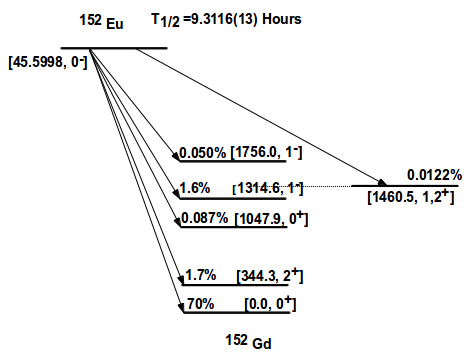}}\label{b3}\hspace*{3.00cm}
\subfloat[]{\label{Figure7:f}\includegraphics[width=5.8cm,height=6.2cm]{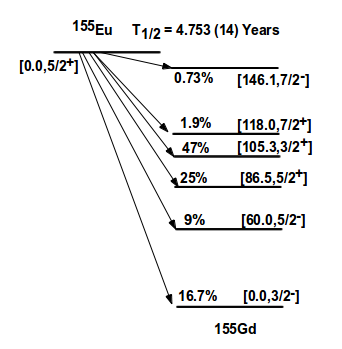}}\label{b4}\\
\caption{$\beta^-$ decay transition ($^{136}Cs$, $^{148}Pm$, $^{152}Eu$, $^{155}Eu$) with neutral atom branchings  are given \cite{nndc}. Here, T$_{1/2}$  is  the total half-life of the  parent  level (including all possible decay channels, viz. $\beta$, $\alpha$, IT etc).  However, only allowed (a), first-forbidden non-unique (nu) and first-forbidden unique (u) $\beta^-$ decay transitions are shown in these figures.}
\label{Figure7}
\end{figure*}
\begin{figure*}
\subfloat[]{\label{Figure8:a}\includegraphics[width=6cm,height=5cm]{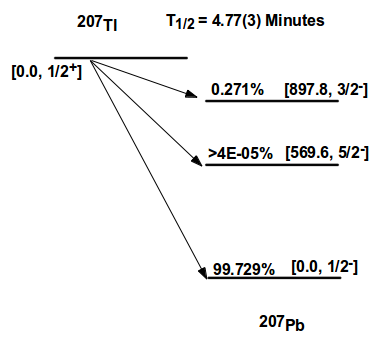}}\label{b5}\hspace*{3.00cm}
\subfloat[]{\label{Figure8:b}\includegraphics[width=6.8cm,height=5.5cm]{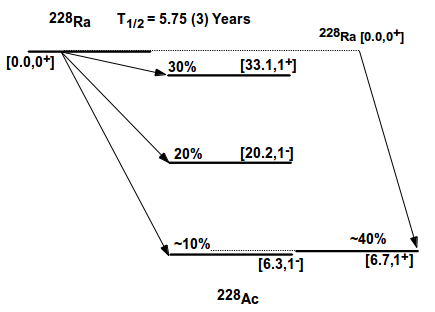}}\label{b6}\\
\vspace{.5cm}
\subfloat[]{\label{Figure8:c}\includegraphics[width=5.8cm,height=5.2cm]{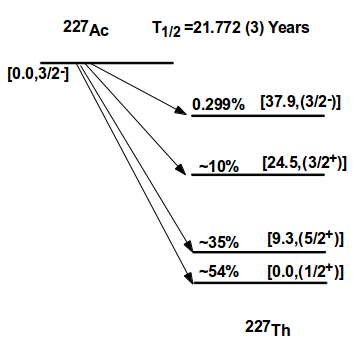}}
\caption{$\beta^-$ decay transition ($^{207}Tl$, $^{228}Ra$, $^{227}Ac$) with neutral atom branchings are given \cite{nndc}. Here, T$_{1/2}$  is  the total half-life of the  parent  level (including all possible decay channels, viz. $\beta$, $\alpha$, IT etc).  However, only allowed (a), first-forbidden non-unique (nu) and first-forbidden unique (u) $\beta^-$ decay transitions are shown in these figures.}
\label{Figure8}
\end{figure*}
Here we present a table containing  $Log~ f_{0}t(f_{1}t)$ values for neutral atoms, bound and continuum state $\beta^-$ decay rates for bare atoms along with the comparison with previous theoretical \cite{takahashi1} as well as  existing data \cite{nndc}, wherever available. Finally a comparative study on bare atom to neutral atom $\beta^{-}$ decay rates for different choices of radii has been presented. 

\vspace{.3cm}

{\bf EXPLANATION OF TABLE} 
\vspace{.2cm}

\underline {Transition Details}:

\vspace{.2cm}

$\bullet$ First column:  $\beta^-$ decay transitions with Neutral atom branchings (FIGS.7 - 9)

$\bullet$ Second column: Participating parent - daughter energy levels  in the transition.

$\bullet$ Third column: Transition types (a, nu and u).

$\bullet$ Forth column:  Neutral atom Q value.

\vspace{.2cm}

\underline {Radii}: 
\vspace{.2cm}

$\bullet$ Fifth column:  $R_1$, $R_2$ and $R_3$ are given in consecutive rows.

\vspace{.2cm}

\underline { $Log~ft$ Calculations}:
\vspace{.2cm}

$\bullet$ Sixth column: $Log~f_0(f_1)$ for different choices of radii (row-wise) calculated using Eq. (2) and (5).

$\bullet$ Seventh column: $Log~f_0t(f_1t)$ for radii $R_1$, $R_2$ and $R_3$  (row-wise), where $Log~ft = Log~f + Log~t$. Here $t$ is the partial half-life of individual transitions (second column) calculated using  T$_{1/2}$ and branching of that particular level (first column).  

$\bullet$ Eighth column:  $Log~f_0t(f_1t)$ values from previous work \cite{takahashi1} (row 1) and existing data \cite{nndc}(row 2).

\vspace{.2cm}

\underline  {Bare Atom Decay Rates}:

\vspace{.2cm}

$\bullet$ Ninth column: Bound state $\beta^-$ decay rates $\lambda_B$ for bare atoms for different choices of radii (row-wise) calculated using Eqs. (1) and (13-15).  

$\bullet$ Tenth column:  $\lambda_B$ from previous work \cite{takahashi1} (row 1).

$\bullet$ Eleventh column: Continuum state $\beta^-$ decay rates $\lambda_C$ for bare atoms for radii $R_1$, $R_2$ and $R_3$  (row-wise) calculated using Eqs. (1) and (10-12).  

$\bullet$ Twelfth column: $\lambda_C$ from previous work \cite{takahashi1} (row 1).

$\bullet$ Thirteenth column: Ratio of bound and continuum state $\beta^-$ decay rates of bare atom for different choices of radii (row-wise).

\vspace{.2cm}

\underline {$\lambda_{Bare} / \lambda_{Neutral}$}:
\vspace{.2cm}

$\bullet$ Fourteenth column: Ratio of  Bare and Neutral atom decay rates  for radii $R_1$, $R_2$ and $R_3$  (row-wise). Here $\lambda_{Bare} = \lambda_B + \lambda_C$.   $\lambda_{Neutral}$ is obtained from  parent level half-life (T$_{1/2}$) and $\beta^-$ branching (first column).
\vspace{.2cm}

\underline{Half-life}:

\vspace{.2cm}

$\bullet$ Fifteenth column: Total $\beta^-$ decay (a, nu, u) half-life of the parent level for  neutral atom, obtained from  T$_{1/2}$ and $\beta^-$ branching (first column).  

 $\bullet$ Sixteenth column:   Total $\beta^-$ decay (a, nu, u) half-life of the parent level for bare atom for radii  $R_1$, $R_2$ and $R_3$  (row-wise). It is obtained by the formula $0.693 \times 1/\sum_i(\lambda_{Bare})_i$, where  {\it i} denotes all the possible a, nu and u type of $\beta^-$ transitions. Here, m: Minutes; h: Hours; d: Days; y : Years.
\vspace{.3cm}

	          \begin{sidewaystable*}

\pagebreak[4]

	     		\vspace*{2.5 cm}
	     		{\label{TABLE -A.I} TABLE A.I : $Log f_{0}t(f_{1}t)$ values , bound and continuum state $\beta^-$ decay rates (bare atom), comparison between neutral atom and bare atom $\beta^{-}$ decay rates for different choices of radii compared with the results of previous theoretical work and experimental data.}
			\vspace*{0.4 cm}

	     		\hspace*{-0.25cm}
	     		\resizebox{22.5cm}{!}{
\begin{tabular}{p{1.5 cm} c c c |c|c c c| c c c c c|c| c c}
	     				\hline\hline	
                \multicolumn{4}{|c|} {Transition Details} & Radii & \multicolumn{3}{|c|} {{$Logf_{0}t (Logf_{1}t)$ Calculations}}  & \multicolumn{5}{|c|} {Bare Atom Decay Rates}  & & & \\ \cline{1-4}\cline{6-8}\cline{9-13}
	     				 &  &  &  & (fm) & \multicolumn{1}{c|} {$Logf_{0} (Logf_{1})$} & \multicolumn{2}{c|} {$Logf_{0}t (Logf_{1}t)$ } &\multicolumn{2}{c|}{ $\lambda_{B}$ (in sec$^{-1}$)}  & \multicolumn{2} {c|}{$\lambda_{C}$ (in sec$^{-1}$)} & $\lambda_{B}/ \lambda_{C}$ & $\lambda_{Bare}/\lambda_{Neutral}$	 &  \multispan{2} Half-life \\  
	     				
	Neutral Atom Branching    & Parent \arrow Daughter                   & Type & $Q_{n}$ &  & \multicolumn{1}{c|}{Present}  & Present & Previous  & Present &  \multicolumn{1}{c|}{Previous} &  Present & \multicolumn{1}{c|}{Previous}  &  &  & Neutral & Bare \\   \cline{6-16}
	 Ref\cite{nndc}  & $[E(keV),J^{\Pi}]$   & & (keV)& $R_1$ & \multicolumn{1}{c|}{(for $R_1$)} &  \multicolumn{1}{c|}{(for $R_1$)}       & Ref\cite{takahashi1}  & \multicolumn{1}{c|}{(for $R_1$)} & \multicolumn{1}{c|}{Ref\cite{takahashi1}} & \multicolumn{1}{c|}{(for $R_1$)}  & \multicolumn{1}{c|} {Ref\cite{takahashi1}}& \multicolumn{1}{c|}{(for $R_1$)} & \multicolumn{1}{c|}{(for $R_1$)} & \multicolumn{1}{c|}{} & {(for $R_1$)} \\ 
                                                 &   & & & $R_2$ & \multicolumn{1}{c|}{(for $R_2$)} &  \multicolumn{1}{c|}{(for $R_2$)}       & Ref\cite{nndc} & \multicolumn{1}{c|}{(for $R_2$)} & \multicolumn{1}{c|}{} & \multicolumn{1}{c|}{(for $R_2$)}  & \multicolumn{1}{c|}{}& \multicolumn{1}{c|}{(for $R_2$)} & \multicolumn{1}{c|}{(for $R_2$)} & \multicolumn{1}{c|}{}  &  {(for $R_2$)} \\
\hline

\end{tabular}}
\end{sidewaystable*}

 \begin{sidewaystable*}
	     		
	     		\vspace*{2.5cm}
	     		~~~~~~~~~TABLE A.I~: (Cotnd.)
			\vspace*{0.3 cm}

	     		\hspace*{-2.00cm}
	     		\resizebox{23.5 cm}{!}{
	     			\begin{tabular}{p{1.5 cm} c c c |c|c c c| c c c c c|c| c c}
	     				\hline\hline	
                \multicolumn{4}{c|} {Transition Details} & Radii & \multicolumn{3}{|c|} {{$Logf_{0}t (Logf_{1}t)$ Calculations}}  & \multicolumn{5}{|c|} {Bare Atom Decay Rates}  & & & \\ \cline{1-4}\cline{6-8}\cline{9-13}
	     				&  &  &  & (fm) & \multicolumn{1}{c|} {$Logf_{0} (Logf_{1})$} & \multicolumn{2}{c|} {$Logf_{0}t (Logf_{1}t)$ } &\multicolumn{2}{c|}{ $\lambda_{B}$ (in sec$^{-1}$)}  & \multicolumn{2} {c|}{$\lambda_{C}$ (in sec$^{-1}$)} & $\lambda_{B}/ \lambda_{C}$ & $\lambda_{Bare}/\lambda_{Neutral}$	 &  \multispan{2} Half-life \\  
	     				
	Neutral Atom Branching    & Parent \arrow Daughter                   & Type & $Q_{n}$ &  & \multicolumn{1}{c|}{Present}  & Present & Previous  & Present &  \multicolumn{1}{c|}{Previous} &  Present & \multicolumn{1}{c|}{Previous}  &  &  & Neutral & Bare \\   \cline{6-16}
Ref\cite{nndc}	   & $[E(keV),J^{\Pi}]$   & & (keV)& $R_1$ & \multicolumn{1}{c|}{(for $R_1$)} &  \multicolumn{1}{c|}{(for $R_1$)}       & Ref\cite{takahashi1}  & \multicolumn{1}{c|}{(for $R_1$)} & \multicolumn{1}{c|}{Ref\cite{takahashi1}} & \multicolumn{1}{c|}{(for $R_1$)}  & \multicolumn{1}{c|} {Ref\cite{takahashi1}}& \multicolumn{1}{c|}{(for $R_1$)} & \multicolumn{1}{c|}{(for $R_1$)} & \multicolumn{1}{c|}{} & {(for $R_1$)} \\ 
                                                 &   & & & $R_2$ & \multicolumn{1}{c|}{(for $R_2$)} &  \multicolumn{1}{c|}{(for $R_2$)}       & Ref\cite{nndc} & \multicolumn{1}{c|}{(for $R_2$)} & \multicolumn{1}{c|}{} & \multicolumn{1}{c|}{(for $R_2$)}  & \multicolumn{1}{c|}{}& \multicolumn{1}{c|}{(for $R_2$)} & \multicolumn{1}{c|}{(for $R_2$)} & \multicolumn{1}{c|}{}  &  {(for $R_2$)} \\
\hline

\end{tabular}}
\end{sidewaystable*}

 \begin{sidewaystable*}
	     		
	     		\vspace*{4.5cm}
	     		~~~~~~~~~TABLE A.I~: (Cotnd.)
			\vspace*{0.3 cm}

	     		\hspace*{-2.00cm}

	     		\resizebox{23.5 cm}{!}{
	     			\begin{tabular}{p{1.5 cm} c c c |c|c c c| c c c c c|c| c c}
	     				\hline\hline	
                \multicolumn{4}{c|} {Transition Details} & Radii & \multicolumn{3}{|c|} {{$Logf_{0}t (Logf_{1}t)$ Calculations}}  & \multicolumn{5}{|c|} {Bare Atom Decay Rates}  & & & \\ \cline{1-4}\cline{6-8}\cline{9-13}
	     				&  &  &  & (fm) & \multicolumn{1}{c|} {$Logf_{0} (Logf_{1})$} & \multicolumn{2}{c|} {$Logf_{0}t (Logf_{1}t)$ } &\multicolumn{2}{c|}{ $\lambda_{B}$ (in sec$^{-1}$)}  & \multicolumn{2} {c|}{$\lambda_{C}$ (in sec$^{-1}$)} & $\lambda_{B}/ \lambda_{C}$ & $\lambda_{Bare}/\lambda_{Neutral}$	 &  \multispan{2} Half-life \\  
	     				
	Neutral Atom Branching    & Parent \arrow Daughter                   & Type & $Q_{n}$ &  & \multicolumn{1}{c|}{Present}  & Present & Previous  & Present &  \multicolumn{1}{c|}{Previous} &  Present & \multicolumn{1}{c|}{Previous}  &  &  & Neutral & Bare \\   \cline{6-16}
Ref\cite{nndc}	   & $[E(keV),J^{\Pi}]$   & & (keV)& $R_1$ & \multicolumn{1}{c|}{(for $R_1$)} &  \multicolumn{1}{c|}{(for $R_1$)}       & Ref\cite{takahashi1}  & \multicolumn{1}{c|}{(for $R_1$)} & \multicolumn{1}{c|}{Ref\cite{takahashi1}} & \multicolumn{1}{c|}{(for $R_1$)}  & \multicolumn{1}{c|} {Ref\cite{takahashi1}}& \multicolumn{1}{c|}{(for $R_1$)} & \multicolumn{1}{c|}{(for $R_1$)} & \multicolumn{1}{c|}{} & {(for $R_1$)} \\ 
                                                 &   & & & $R_2$ & \multicolumn{1}{c|}{(for $R_2$)} &  \multicolumn{1}{c|}{(for $R_2$)}       & Ref\cite{nndc} & \multicolumn{1}{c|}{(for $R_2$)} & \multicolumn{1}{c|}{} & \multicolumn{1}{c|}{(for $R_2$)}  & \multicolumn{1}{c|}{}& \multicolumn{1}{c|}{(for $R_2$)} & \multicolumn{1}{c|}{(for $R_2$)} & \multicolumn{1}{c|}{}  &  {(for $R_2$)} \\
\hline	

\end{tabular}}
\end{sidewaystable*}

 \begin{sidewaystable*}
	     		
	     		\vspace*{6.5 cm}

	     		~~~~~~~~~TABLE A.I~: (Cotnd.)
			\vspace*{0.3 cm}

	     		\hspace*{-2.00cm}
	     		\resizebox{22.0 cm}{!}{
	     			\begin{tabular}{p{1.5 cm} c c c |c|c c c| c c c c c|c| c c}
	     				\hline\hline	
                \multicolumn{4}{c|} {Transition Details} & Radii & \multicolumn{3}{|c|} {{$Logf_{0}t (Logf_{1}t)$ Calculations}}  & \multicolumn{5}{|c|} {Bare Atom Decay Rates}  & & & \\ \cline{1-4}\cline{6-8}\cline{9-13}
	     				&  &  &  & (fm) & \multicolumn{1}{c|} {$Logf_{0} (Logf_{1})$} & \multicolumn{2}{c|} {$Logf_{0}t (Logf_{1}t)$ } &\multicolumn{2}{c|}{ $\lambda_{B}$ (in sec$^{-1}$)}  & \multicolumn{2} {c|}{$\lambda_{C}$ (in sec$^{-1}$)} & $\lambda_{B}/ \lambda_{C}$ & $\lambda_{Bare}/\lambda_{Neutral}$	 &  \multispan{2} Half-life \\  
	     				
	Neutral Atom Branching    & Parent \arrow Daughter                   & Type & $Q_{n}$ &  & \multicolumn{1}{c|}{Present}  & Present & Previous  & Present &  \multicolumn{1}{c|}{Previous} &  Present & \multicolumn{1}{c|}{Previous}  &  &  & Neutral & Bare \\   \cline{6-16}
	Ref\cite{nndc}   & $[E(keV),J^{\Pi}]$   & & (keV)& $R_1$ & \multicolumn{1}{c|}{(for $R_1$)} &  \multicolumn{1}{c|}{(for $R_1$)}       & Ref\cite{takahashi1}  & \multicolumn{1}{c|}{(for $R_1$)} & \multicolumn{1}{c|}{Ref\cite{takahashi1}} & \multicolumn{1}{c|}{(for $R_1$)}  & \multicolumn{1}{c|} {Ref\cite{takahashi1}}& \multicolumn{1}{c|}{(for $R_1$)} & \multicolumn{1}{c|}{(for $R_1$)} & \multicolumn{1}{c|}{} & {(for $R_1$)} \\ 
                                                 &   & & & $R_2$ & \multicolumn{1}{c|}{(for $R_2$)} &  \multicolumn{1}{c|}{(for $R_2$)}       & Ref\cite{nndc} & \multicolumn{1}{c|}{(for $R_2$)} & \multicolumn{1}{c|}{} & \multicolumn{1}{c|}{(for $R_2$)}  & \multicolumn{1}{c|}{}& \multicolumn{1}{c|}{(for $R_2$)} & \multicolumn{1}{c|}{(for $R_2$)} & \multicolumn{1}{c|}{}  &  {(for $R_2$)} \\
\hline
\end{tabular}}
\rule{0pt} {1 cm}

$\ddagger$ Not calculated here, see Appendix \ref{spin} for details.
\end{sidewaystable*}

	    	   
	       \begin{sidewaystable*}

	     		
			\vspace*{3.5 cm}
			~~~~~~~~~TABLE A.I~: (Cotnd.)
			\vspace*{0.3 cm}

	     		\hspace*{-2.00cm}
	     		\resizebox{22.5 cm}{!}{
	     			\begin{tabular}{p{1.5 cm} c c c |c|c c c| c c c c c|c| c c}
	     				\hline\hline	
                \multicolumn{4}{c|} {Transition Details} & Radii & \multicolumn{3}{|c|} {{$Logf_{0}t (Logf_{1}t)$ Calculations}}  & \multicolumn{5}{|c|} {Bare Atom Decay Rates}  & & & \\ \cline{1-4}\cline{6-8}\cline{9-13}
	     				&  &  &  & (fm) & \multicolumn{1}{c|} {$Logf_{0} (Logf_{1})$} & \multicolumn{2}{c|} {$Logf_{0}t (Logf_{1}t)$ } &\multicolumn{2}{c|}{ $\lambda_{B}$ (in sec$^{-1}$)}  & \multicolumn{2} {c|}{$\lambda_{C}$ (in sec$^{-1}$)} & $\lambda_{B}/ \lambda_{C}$ & $\lambda_{Bare}/\lambda_{Neutral}$	 &  \multispan{2} Half-life \\  
	     				
	Neutral Atom Branching    & Parent \arrow Daughter                   & Type & $Q_{n}$ &  & \multicolumn{1}{c|}{Present}  & Present & Previous  & Present &  \multicolumn{1}{c|}{Previous} &  Present & \multicolumn{1}{c|}{Previous}  &  &  & Neutral & Bare \\   \cline{6-16}
	Ref\cite{nndc}   & $[E(keV),J^{\Pi}]$   & & (keV)& $R_1$ & \multicolumn{1}{c|}{(for $R_1$)} &  \multicolumn{1}{c|}{(for $R_1$)}       & Ref\cite{takahashi1}  & \multicolumn{1}{c|}{(for $R_1$)} & \multicolumn{1}{c|}{Ref\cite{takahashi1}} & \multicolumn{1}{c|}{(for $R_1$)}  & \multicolumn{1}{c|} {Ref\cite{takahashi1}}& \multicolumn{1}{c|}{(for $R_1$)} & \multicolumn{1}{c|}{(for $R_1$)} & \multicolumn{1}{c|}{} & {(for $R_1$)} \\ 
                                                 &   & & & $R_2$ & \multicolumn{1}{c|}{(for $R_2$)} &  \multicolumn{1}{c|}{(for $R_2$)}       & Ref\cite{nndc} & \multicolumn{1}{c|}{(for $R_2$)} & \multicolumn{1}{c|}{} & \multicolumn{1}{c|}{(for $R_2$)}  & \multicolumn{1}{c|}{}& \multicolumn{1}{c|}{(for $R_2$)} & \multicolumn{1}{c|}{(for $R_2$)} & \multicolumn{1}{c|}{}  &  {(for $R_2$)} \\
\hline

\end{tabular}}

\end{sidewaystable*}

	    	   
	        \begin{sidewaystable*}
	     		
	     		\vspace*{6.5 cm}

	     		~~~~~~~~~TABLE A.I~: (Cotnd.)
			\vspace*{0.3 cm}

	     		\hspace*{-2.00cm}
	     		\resizebox{24.5 cm}{!}{
	     			\begin{tabular}{p{1.5 cm} c c c |c|c c c| c c c c c|c| c c}
	     				\hline\hline	
                \multicolumn{4}{c|} {Transition Details} & Radii & \multicolumn{3}{|c|} {{$Logf_{0}t (Logf_{1}t)$ Calculations}}  & \multicolumn{5}{|c|} {Bare Atom Decay Rates}  & & & \\ \cline{1-4}\cline{6-8}\cline{9-13}
	     				&  &  &  & (fm) & \multicolumn{1}{c|} {$Logf_{0} (Logf_{1})$} & \multicolumn{2}{c|} {$Logf_{0}t (Logf_{1}t)$ } &\multicolumn{2}{c|}{ $\lambda_{B}$ (in sec$^{-1}$)}  & \multicolumn{2} {c|}{$\lambda_{C}$ (in sec$^{-1}$)} & $\lambda_{B}/ \lambda_{C}$ & $\lambda_{Bare}/\lambda_{Neutral}$	 &  \multispan{2} Half-life \\  
	     				
	Neutral Atom Branching    & Parent \arrow Daughter                   & Type & $Q_{n}$ &  & \multicolumn{1}{c|}{Present}  & Present & Previous  & Present &  \multicolumn{1}{c|}{Previous} &  Present & \multicolumn{1}{c|}{Previous}  &  &  & Neutral & Bare \\   \cline{6-16}
	Ref\cite{nndc}   & $[E(keV),J^{\Pi}]$   & & (keV)& $R_1$ & \multicolumn{1}{c|}{(for $R_1$)} &  \multicolumn{1}{c|}{(for $R_1$)}       & Ref\cite{takahashi1}  & \multicolumn{1}{c|}{(for $R_1$)} & \multicolumn{1}{c|}{Ref\cite{takahashi1}} & \multicolumn{1}{c|}{(for $R_1$)}  & \multicolumn{1}{c|} {Ref\cite{takahashi1}}& \multicolumn{1}{c|}{(for $R_1$)} & \multicolumn{1}{c|}{(for $R_1$)} & \multicolumn{1}{c|}{} & {(for $R_1$)} \\ 
                                                 &   & & & $R_2$ & \multicolumn{1}{c|}{(for $R_2$)} &  \multicolumn{1}{c|}{(for $R_2$)}       & Ref\cite{nndc} & \multicolumn{1}{c|}{(for $R_2$)} & \multicolumn{1}{c|}{} & \multicolumn{1}{c|}{(for $R_2$)}  & \multicolumn{1}{c|}{}& \multicolumn{1}{c|}{(for $R_2$)} & \multicolumn{1}{c|}{(for $R_2$)} & \multicolumn{1}{c|}{}  &  {(for $R_2$)} \\
\hline

\end{tabular}}
\end{sidewaystable*}

	    	   
	       \begin{sidewaystable*}

	     		
			\vspace*{5.5 cm}

			~~~~~~~~~TABLE A.I~: (Cotnd.)
			\vspace*{0.3 cm}

	     		\hspace*{-2.00cm}
	     		\resizebox{22.5 cm}{!}{
	     			\begin{tabular}{p{1.5 cm} c c c |c|c c c| c c c c c|c| c c}
	     				\hline\hline	
                \multicolumn{4}{c|} {Transition Details} & Radii & \multicolumn{3}{|c|} {{$Logf_{0}t (Logf_{1}t)$ Calculations}}  & \multicolumn{5}{|c|} {Bare Atom Decay Rates}  & & & \\ \cline{1-4}\cline{6-8}\cline{9-13}
	     				&  &  &  & (fm) & \multicolumn{1}{c|} {$Logf_{0} (Logf_{1})$} & \multicolumn{2}{c|} {$Logf_{0}t (Logf_{1}t)$ } &\multicolumn{2}{c|}{ $\lambda_{B}$ (in sec$^{-1}$)}  & \multicolumn{2} {c|}{$\lambda_{C}$ (in sec$^{-1}$)} & $\lambda_{B}/ \lambda_{C}$ & $\lambda_{Bare}/\lambda_{Neutral}$	 &  \multispan{2} Half-life \\  
	     				
	Neutral Atom Branching    & Parent \arrow Daughter                   & Type & $Q_{n}$ &  & \multicolumn{1}{c|}{Present}  & Present & Previous  & Present &  \multicolumn{1}{c|}{Previous} &  Present & \multicolumn{1}{c|}{Previous}  &  &  & Neutral & Bare \\   \cline{6-16}
	Ref\cite{nndc}   & $[E(keV),J^{\Pi}]$   & & (keV)& $R_1$ & \multicolumn{1}{c|}{(for $R_1$)} &  \multicolumn{1}{c|}{(for $R_1$)}       & Ref\cite{takahashi1}  & \multicolumn{1}{c|}{(for $R_1$)} & \multicolumn{1}{c|}{Ref\cite{takahashi1}} & \multicolumn{1}{c|}{(for $R_1$)}  & \multicolumn{1}{c|} {Ref\cite{takahashi1}}& \multicolumn{1}{c|}{(for $R_1$)} & \multicolumn{1}{c|}{(for $R_1$)} & \multicolumn{1}{c|}{} & {(for $R_1$)} \\ 
                                                 &   & & & $R_2$ & \multicolumn{1}{c|}{(for $R_2$)} &  \multicolumn{1}{c|}{(for $R_2$)}       & Ref\cite{nndc} & \multicolumn{1}{c|}{(for $R_2$)} & \multicolumn{1}{c|}{} & \multicolumn{1}{c|}{(for $R_2$)}  & \multicolumn{1}{c|}{}& \multicolumn{1}{c|}{(for $R_2$)} & \multicolumn{1}{c|}{(for $R_2$)} & \multicolumn{1}{c|}{}  &  {(for $R_2$)} \\
\hline

\end{tabular}}
\end{sidewaystable*}

	    	   
	        \begin{sidewaystable*}
	     		
	     		\vspace*{3.5 cm}

	     		~~~~~~~~~TABLE A.I~: (Cotnd.)
			\vspace*{0.3 cm}

	     		\hspace*{-2.00cm}
	     		\resizebox{23.5 cm}{!}{
	     			\begin{tabular}{p{1.5 cm} c c c |c|c c c| c c c c c|c| c c}
	     				\hline\hline	
                \multicolumn{4}{c|} {Transition Details} & Radii & \multicolumn{3}{|c|} {{$Logf_{0}t (Logf_{1}t)$ Calculations}}  & \multicolumn{5}{|c|} {Bare Atom Decay Rates}  & & & \\ \cline{1-4}\cline{6-8}\cline{9-13}
	     				&  &  &  & (fm) & \multicolumn{1}{c|} {$Logf_{0} (Logf_{1})$} & \multicolumn{2}{c|} {$Logf_{0}t (Logf_{1}t)$ } &\multicolumn{2}{c|}{ $\lambda_{B}$ (in sec$^{-1}$)}  & \multicolumn{2} {c|}{$\lambda_{C}$ (in sec$^{-1}$)} & $\lambda_{B}/ \lambda_{C}$ & $\lambda_{Bare}/\lambda_{Neutral}$	 &  \multispan{2} Half-life \\  
	     				
	Neutral Atom Branching    & Parent \arrow Daughter                   & Type & $Q_{n}$ &  & \multicolumn{1}{c|}{Present}  & Present & Previous  & Present &  \multicolumn{1}{c|}{Previous} &  Present & \multicolumn{1}{c|}{Previous}  &  &  & Neutral & Bare \\   \cline{6-16}
	Ref\cite{nndc}   & $[E(keV),J^{\Pi}]$   & & (keV)& $R_1$ & \multicolumn{1}{c|}{(for $R_1$)} &  \multicolumn{1}{c|}{(for $R_1$)}       & Ref\cite{takahashi1}  & \multicolumn{1}{c|}{(for $R_1$)} & \multicolumn{1}{c|}{Ref\cite{takahashi1}} & \multicolumn{1}{c|}{(for $R_1$)}  & \multicolumn{1}{c|} {Ref\cite{takahashi1}}& \multicolumn{1}{c|}{(for $R_1$)} & \multicolumn{1}{c|}{(for $R_1$)} & \multicolumn{1}{c|}{} & {(for $R_1$)} \\ 
                                                 &   & & & $R_2$ & \multicolumn{1}{c|}{(for $R_2$)} &  \multicolumn{1}{c|}{(for $R_2$)}       & Ref\cite{nndc} & \multicolumn{1}{c|}{(for $R_2$)} & \multicolumn{1}{c|}{} & \multicolumn{1}{c|}{(for $R_2$)}  & \multicolumn{1}{c|}{}& \multicolumn{1}{c|}{(for $R_2$)} & \multicolumn{1}{c|}{(for $R_2$)} & \multicolumn{1}{c|}{}  &  {(for $R_2$)} \\
\hline	

\end{tabular}}
\end{sidewaystable*}
	    
	    	   
	        \begin{sidewaystable*}
	     		
	     		\vspace*{6.5 cm}

	     		~~~~~~~~~TABLE A.I~: (Cotnd.)
			\vspace*{0.3 cm}

	     		\hspace*{-2.5cm}
	     		\resizebox{22.5 cm}{!}{
	     			\begin{tabular}{p{1.5 cm} c c c |c|c c c| c c c c c|c| c c}
	     				\hline\hline	
                \multicolumn{4}{c|} {Transition Details} & Radii & \multicolumn{3}{|c|} {{$Logf_{0}t (Logf_{1}t)$ Calculations}}  & \multicolumn{5}{|c|} {Bare Atom Decay Rates}  & & & \\ \cline{1-4}\cline{6-8}\cline{9-13}
	     				&  &  &  & (fm) & \multicolumn{1}{c|} {$Logf_{0} (Logf_{1})$} & \multicolumn{2}{c|} {$Logf_{0}t (Logf_{1}t)$ } &\multicolumn{2}{c|}{ $\lambda_{B}$ (in sec$^{-1}$)}  & \multicolumn{2} {c|}{$\lambda_{C}$ (in sec$^{-1}$)} & $\lambda_{B}/ \lambda_{C}$ & $\lambda_{Bare}/\lambda_{Neutral}$	 &  \multispan{2} Half-life \\  
	     				
	Neutral Atom Branching    & Parent \arrow Daughter                   & Type & $Q_{n}$ &  & \multicolumn{1}{c|}{Present}  & Present & Previous  & Present &  \multicolumn{1}{c|}{Previous} &  Present & \multicolumn{1}{c|}{Previous}  &  &  & Neutral & Bare \\   \cline{6-16}
	Ref\cite{nndc}   & $[E(keV),J^{\Pi}]$   & & (keV)& $R_1$ & \multicolumn{1}{c|}{(for $R_1$)} &  \multicolumn{1}{c|}{(for $R_1$)}       & Ref\cite{takahashi1}  & \multicolumn{1}{c|}{(for $R_1$)} & \multicolumn{1}{c|}{Ref\cite{takahashi1}} & \multicolumn{1}{c|}{(for $R_1$)}  & \multicolumn{1}{c|} {Ref\cite{takahashi1}}& \multicolumn{1}{c|}{(for $R_1$)} & \multicolumn{1}{c|}{(for $R_1$)} & \multicolumn{1}{c|}{} & {(for $R_1$)} \\ 
                                                 &   & & & $R_2$ & \multicolumn{1}{c|}{(for $R_2$)} &  \multicolumn{1}{c|}{(for $R_2$)}       & Ref\cite{nndc} & \multicolumn{1}{c|}{(for $R_2$)} & \multicolumn{1}{c|}{} & \multicolumn{1}{c|}{(for $R_2$)}  & \multicolumn{1}{c|}{}& \multicolumn{1}{c|}{(for $R_2$)} & \multicolumn{1}{c|}{(for $R_2$)} & \multicolumn{1}{c|}{}  &  {(for $R_2$)} \\
\hline

\end{tabular}}
\end{sidewaystable*}

	    	     
	        \begin{sidewaystable*}
	     		
	     		\vspace*{7.5 cm}

	     		~~~~~~~~~TABLE A.I~: (Cotnd.)
			\vspace*{0.3 cm}

	     		\hspace*{-2.00cm}
	     		\resizebox{21.0 cm}{!}{
	     			\begin{tabular}{p{1.5 cm} c c c |c|c c c| c c c c c|c| c c}
	     				\hline\hline	
                \multicolumn{4}{c|} {Transition Details} & Radii & \multicolumn{3}{|c|} {{$Logf_{0}t (Logf_{1}t)$ Calculations}}  & \multicolumn{5}{|c|} {Bare Atom Decay Rates}  & & & \\ \cline{1-4}\cline{6-8}\cline{9-13}
	     				&  &  &  & (fm) & \multicolumn{1}{c|} {$Logf_{0} (Logf_{1})$} & \multicolumn{2}{c|} {$Logf_{0}t (Logf_{1}t)$ } &\multicolumn{2}{c|}{ $\lambda_{B}$ (in sec$^{-1}$)}  & \multicolumn{2} {c|}{$\lambda_{C}$ (in sec$^{-1}$)} & $\lambda_{B}/ \lambda_{C}$ & $\lambda_{Bare}/\lambda_{Neutral}$	 &  \multispan{2} Half-life \\  
	     				
	Neutral Atom Branching    & Parent \arrow Daughter                   & Type & $Q_{n}$ &  & \multicolumn{1}{c|}{Present}  & Present & Previous  & Present &  \multicolumn{1}{c|}{Previous} &  Present & \multicolumn{1}{c|}{Previous}  &  &  & Neutral & Bare \\   \cline{6-16}
	Ref\cite{nndc}   & $[E(keV),J^{\Pi}]$   & & (keV)& $R_1$ & \multicolumn{1}{c|}{(for $R_1$)} &  \multicolumn{1}{c|}{(for $R_1$)}       & Ref\cite{takahashi1}  & \multicolumn{1}{c|}{(for $R_1$)} & \multicolumn{1}{c|}{Ref\cite{takahashi1}} & \multicolumn{1}{c|}{(for $R_1$)}  & \multicolumn{1}{c|} {Ref\cite{takahashi1}}& \multicolumn{1}{c|}{(for $R_1$)} & \multicolumn{1}{c|}{(for $R_1$)} & \multicolumn{1}{c|}{} & {(for $R_1$)} \\ 
                                                 &   & & & $R_2$ & \multicolumn{1}{c|}{(for $R_2$)} &  \multicolumn{1}{c|}{(for $R_2$)}       & Ref\cite{nndc} & \multicolumn{1}{c|}{(for $R_2$)} & \multicolumn{1}{c|}{} & \multicolumn{1}{c|}{(for $R_2$)}  & \multicolumn{1}{c|}{}& \multicolumn{1}{c|}{(for $R_2$)} & \multicolumn{1}{c|}{(for $R_2$)} & \multicolumn{1}{c|}{}  &  {(for $R_2$)} \\
\hline

\end{tabular}}
\rule{0pt} {1 cm}

$\ddagger$ Not calculated here, see Appendix \ref{spin} for details.
\end{sidewaystable*}

	    	     
	        \begin{sidewaystable*}
	     		
	     		\vspace*{7.5 cm}

	     		~~~~~~~~~TABLE A.I~: (Cotnd.)
			\vspace*{0.3 cm}

	     		\hspace*{-2.00cm}
	     		\resizebox{21.0 cm}{!}{
	     			\begin{tabular}{p{1.5 cm} c c c |c|c c c| c c c c c|c| c c}
	     				\hline\hline	
                \multicolumn{4}{c|} {Transition Details} & Radii & \multicolumn{3}{|c|} {{$Logf_{0}t (Logf_{1}t)$ Calculations}}  & \multicolumn{5}{|c|} {Bare Atom Decay Rates}  & & & \\ \cline{1-4}\cline{6-8}\cline{9-13}
	     				&  &  &  & (fm) & \multicolumn{1}{c|} {$Logf_{0} (Logf_{1})$} & \multicolumn{2}{c|} {$Logf_{0}t (Logf_{1}t)$ } &\multicolumn{2}{c|}{ $\lambda_{B}$ (in sec$^{-1}$)}  & \multicolumn{2} {c|}{$\lambda_{C}$ (in sec$^{-1}$)} & $\lambda_{B}/ \lambda_{C}$ & $\lambda_{Bare}/\lambda_{Neutral}$	 &  \multispan{2} Half-life \\  
	     				
	Neutral Atom Branching    & Parent \arrow Daughter                   & Type & $Q_{n}$ &  & \multicolumn{1}{c|}{Present}  & Present & Previous  & Present &  \multicolumn{1}{c|}{Previous} &  Present & \multicolumn{1}{c|}{Previous}  &  &  & Neutral & Bare \\   \cline{6-16}
	Ref\cite{nndc}   & $[E(keV),J^{\Pi}]$   & & (keV)& $R_1$ & \multicolumn{1}{c|}{(for $R_1$)} &  \multicolumn{1}{c|}{(for $R_1$)}       & Ref\cite{takahashi1}  & \multicolumn{1}{c|}{(for $R_1$)} & \multicolumn{1}{c|}{Ref\cite{takahashi1}} & \multicolumn{1}{c|}{(for $R_1$)}  & \multicolumn{1}{c|} {Ref\cite{takahashi1}}& \multicolumn{1}{c|}{(for $R_1$)} & \multicolumn{1}{c|}{(for $R_1$)} & \multicolumn{1}{c|}{} & {(for $R_1$)} \\ 
                                                 &   & & & $R_2$ & \multicolumn{1}{c|}{(for $R_2$)} &  \multicolumn{1}{c|}{(for $R_2$)}       & Ref\cite{nndc} & \multicolumn{1}{c|}{(for $R_2$)} & \multicolumn{1}{c|}{} & \multicolumn{1}{c|}{(for $R_2$)}  & \multicolumn{1}{c|}{}& \multicolumn{1}{c|}{(for $R_2$)} & \multicolumn{1}{c|}{(for $R_2$)} & \multicolumn{1}{c|}{}  &  {(for $R_2$)} \\
\hline

\end{tabular}}
\rule{0pt} {1 cm}


* The mismatch between $\lambda_{B}$ value may arise from the typographical error in the tabulation of $\lambda_{B}$ in Ref. \cite{takahashi1}.

\end{sidewaystable*}


	          \begin{sidewaystable*}
	     		
	     		\vspace*{7.5 cm}
	     		~~~~~~~~~TABLE A.I~: (Cotnd.)
			\vspace*{0.3 cm}

	     		\hspace*{-1.00cm}
	     		\resizebox{21.0 cm}{!}{
	     			\begin{tabular}{c| c c c |c|c c c| c c c c c|c| c c}
	     				\hline\hline	
                \multicolumn{4}{c|} {Transition Details} & Radii & \multicolumn{3}{|c|} {{$Logf_{0}t (Logf_{1}t)$ Calculations}}  & \multicolumn{5}{|c|} {Bare Atom Decay Rates}  & & & \\ \cline{1-4}\cline{6-8}\cline{9-13}
	     				&  &  &  & (fm) & \multicolumn{1}{c|} {$Logf_{0} (Logf_{1})$} & \multicolumn{2}{c|} {$Logf_{0}t (Logf_{1}t)$ } &\multicolumn{2}{c|}{ $\lambda_{B}$ (in sec$^{-1}$)}  & \multicolumn{2} {c|}{$\lambda_{C}$ (in sec$^{-1}$)} & $\lambda_{B}/ \lambda_{C}$ & $\lambda_{Bare}/\lambda_{Neutral}$	 &  \multispan{2} Half-life \\  
	Neutral      & Parent \arrow Daughter                   & Type & $Q_{n}$ &  & \multicolumn{1}{c|}{Present}  & Present & Previous  & Present &  \multicolumn{1}{c|}{Previous} &  Present & \multicolumn{1}{c|}{Previous}  &  &  & Neutral & Bare \\   \cline{6-16}
	 Atom    & $[E(keV),J^{\Pi}]$   & & (keV)& $R_1$ & \multicolumn{1}{c|}{(for $R_1$)} &  \multicolumn{1}{c|}{(for $R_1$)}       & Ref\cite{takahashi1}  & \multicolumn{1}{c|}{(for $R_1$)} & \multicolumn{1}{c|}{Ref\cite{takahashi1}} & \multicolumn{1}{c|}{(for $R_1$)}  & \multicolumn{1}{c|} {Ref\cite{takahashi1}}& \multicolumn{1}{c|}{(for $R_1$)} & \multicolumn{1}{c|}{(for $R_1$)} & \multicolumn{1}{c|}{} & {(for $R_1$)} \\ 
           Branching                                      &   & & & $R_2$ & \multicolumn{1}{c|}{(for $R_2$)} &  \multicolumn{1}{c|}{(for $R_2$)}       & Ref\cite{nndc} & \multicolumn{1}{c|}{(for $R_2$)} & \multicolumn{1}{c|}{} & \multicolumn{1}{c|}{(for $R_2$)}  & \multicolumn{1}{c|}{}& \multicolumn{1}{c|}{(for $R_2$)} & \multicolumn{1}{c|}{(for $R_2$)} & \multicolumn{1}{c|}{}  &  {(for $R_2$)} \\
  Ref\cite{nndc} &   & & & $R_3$ & \multicolumn{1}{c|}{(for $R_3$)} &  \multicolumn{1}{c|}{(for $R_3$)}       &  & \multicolumn{1}{c|}{(for $R_3$)} & \multicolumn{1}{c|}{ } &  \multicolumn{1}{c|}{(for $R_3$)} & \multicolumn{1}{c|}{}& \multicolumn{1}{c|}{(for $R_3$)} & \multicolumn{1}{c|}{(for $R_3$)} & \multicolumn{1}{c|}{} & {(for $R_3$)}  \\ \hline


& $^{207}Tl$ \arrow $^{207}Pb$                                                                  &     &       & 7.099 & 2.6681  & 5.126 & -  & $4.10\times{10^{-4}}$ & * &  $2.53\times{10^{-3}}$  & * & $1.62\times{10^{-1}}$ &  $1.22\times{10^{0}}$  &  &   \\  

& $\left[0.0  , \dfrac{1}{2}^{+}\right]$    $\left[0.0 ,\dfrac{1}{2}^{-}\right]$         & nu & 1418.000 & 5.494 & 2.7115 & 5.170 & 5.108 & $4.12\times{10^{-4}}$  &  & $2.53\times{10^{-3}}$  &  & $1.63\times{10^{-1}}$ & $1.22\times{10^{0}}$   &  &  \\ 
	   
                                                                                  &          &    &      & 6.484  & 2.6780  & 5.136 &     & $4.16\times{10^{-4}}$  &  & $2.53\times{10^{-3}}$  &  & $1.65\times{10^{-1}}$ & $1.22\times{10^{0}}$   &  &  \\ 	   \cline{2-14}

& $^{207}Tl$ \arrow $^{207}Pb$                                                                  &     &       & 7.099 & 2.2681 & 11.123 &  -   & $5.08\times{10^{-10}}$ & - &  $9.16\times{10^{-10}}$  & - & $5.54\times{10^{-1}}$ & $1.47\times{10^{0}}$    &  & 3.91 m  \\  

FIG. \ref{Figure8:a} & $\left[0.0  , \dfrac{1}{2}^{+}\right]$    $\left[569.6 ,\dfrac{5}{2}^{-}\right]$         & u & 848.800 & 5.494 & 2.3013 & 11.156  & $>$10.5 & $5.24\times{10^{-10}}$  &  & $9.17\times{10^{-10}}$  &  & $5.71\times{10^{-1}}$ & $1.49\times{10^{0}}$    & 4.77 m & 3.92 m \\ 
	   
                                                                                    &        &    &      & 6.484 & 2.2756  & 11.131 &     & $5.18\times{10^{-10}}$  &  & $9.15\times{10^{-10}}$  &  & $5.66\times{10^{-1}}$ &  $1.48\times{10^{0}}$  &  & 3.91 m \\ 	   \cline{2-14}

& $^{207}Tl$ \arrow $^{207}Pb$                                                                  &     &       & 7.099  & 1.1450 & 6.169 &  -   & $5.83\times{10^{-6}}$ & - &  $6.40\times{10^{-6}}$  & - & $9.10\times{10^{-1}}$ &  $1.86\times{10^{0}}$  &   &   \\  

& $\left[0.0  , \dfrac{1}{2}^{+}\right]$    $\left[897.8 ,\dfrac{3}{2}^{-}\right]$         & nu & 520.200 & 5.494 & 1.1884  & 6.212 & 6.157 & $5.88\times{10^{-6}}$  &  & $6.41\times{10^{-6}}$  &  & $9.17\times{10^{-1}}$ & $1.87\times{10^{0}}$   &  &  \\ 
	   
                                                                                   &         &    &      & 6.484  & 1.1549 & 6.179 &     & $5.92\times{10^{-6}}$  &  & $6.40\times{10^{-6}}$  &  & $9.24\times{10^{-1}}$ & $1.88\times{10^{0}}$   &  &  \\ 	   \hline

& $^{228}Ra$ \arrow $^{228}Ac$                                                                  &     &       & 7.331 & -2.1848 & 7.074 &  -   & $5.90\times{10^{-8}}$ & - & $5.03\times{10^{-11}}$  & - & $1.17\times{10^{3}}$ & $1.55\times{10^{2}}$  &  &   \\  

& $\left[0.0  , 0^{+}\right]$    $\left[6.3 , 1^{-}\right]$         & nu & 39.500 & - & - & - & $\sim$ 7.1 & -  &  &-  &  & - &  -  &   &  \\ 
	   
                                                                                     &       &    &      & 6.707 & -2.1616  & 7.097 &     & $ 5.85\times{10^{-8}}$  &  & $5.04\times{10^{-11}}$  &  & $1.16\times{10^{3}}$ & $1.53\times{10^{2}}$   &   &   \\ 	  \cline{2-14}

& $^{228}Ra$ \arrow $^{228}Ac$                                                                  &     &       & 7.331 & -2.1848 & 6.472 & (6.5)    & $2.35\times{10^{-7}}$ & $2.3\times{10^{-7}}$ & $ 1.87\times{10^{-10}}$  & $1.8\times{10^{-10}}$ & $1.26\times{10^{3}}$ & $1.54\times{10^{2}}$  &  &  1.98 d \\  

& $\left[0.0  , 0^{+}\right]$    $\left[6.7 , 1^{+}\right]$         & a & 39.100 & - & - & -& $\sim$ 6.5 & -  &  &-  &  & - &  -  &  5.75 y & - \\ 
	   
                                                                                   &         &    &      & 6.707 & -2.1616 & 6.495 &     & $2.33\times{10^{-7}}$  &  & $1.86\times{10^{-10}}$  &  & $1.25\times{10^{3}}$ &  $1.52\times{10^{2}}$  &  & 2.00 d  \\ 	  \cline{2-14}

FIG. \ref{Figure8:b} & $^{228}Ra$ \arrow $^{228}Ac$                                                                  &     &       & 7.331 & -2.7515 & 6.206 & -    & $3.50\times{10^{-7}}$ & - & $ 8.97\times{10^{-12}}$  & - & $3.94\times{10^{4}}$ & $4.58\times{10^{2}}$  &  &   \\  

& $\left[0.0  , 0^{+}\right]$    $\left[20.2 , 1^{-}\right]$         & nu & 25.600 & - & - & -&  6.20 & -  &  &-  &  & - &  -  &   &  \\ 
	   
                                                                                  &          &    &      & 6.707 & -2.7284 & 6.229 &     & $3.47\times{10^{-7}}$  &  & $8.96\times{10^{-12}}$  &  & $3.88\times{10^{4}}$ &  $4.55\times{10^{2}}$  &   &   \\ 	  \cline{2-14}	

& $^{228}Ra$ \arrow $^{228}Ac$                                                                  &     &       & 7.331  & -3.6585  & 5.123 & (5.0)    & $3.40\times{10^{-6}}$ & $4.9\times{10^{-6}}$ &  0  & 0  & $\infty$ & $2.96\times{10^{3}}$   &  &  \\  

& $\left[0.0  , 0^{+}\right]$    $\left[33.1 , 1^{+}\right]$         & a & 12.700 & - & - & -&  5.12 & -  &  &-  &  & - & -   &   &  \\ 
	   
                                                                                 &           &    &      & 6.707  & -3.6353& 5.146 &     & $3.37\times{10^{-6}}$  &  & 0  &  & $\infty$ &    $2.94\times{10^{3}}$ &   &   \\ 	  \hline	


 & $^{227}Ac$ \arrow $^{227}Th$                                                                  &     &       & 7.320 & -1.9817  & 7.123 & 7.09  & $6.44\times{10^{-8}}$ & $7.2\times{10^{-8}}$ &  $1.04\times{10^{-10}}$  & $9.8\times{10^{-11}}$ & $6.20\times{10^{2}}$ &  $1.18\times{10^{2}}$  &  &   \\  

& $\left[0.0  , \dfrac{3}{2}^{-}\right]$    $\left[0.0 ,\left(\dfrac{1}{2}^{+}\right)\right]$         & nu & 44.800 & 5.740 & -1.9311 & 7.173 & $\sim$7.1 & $6.50\times{10^{-8}}$  &  & $1.04\times{10^{-10}}$  &  & $6.27\times{10^{2}}$ &  $1.20\times{10^{2}}$  &  &  \\ 
	   
                                                                                    &        &    &      & 6.696  & -1.9579  & 7.146 &     & $6.39\times{10^{-8}}$  &  & $1.04\times{10^{-10}}$  &  & $6.15\times{10^{2}}$ &  $1.17\times{10^{2}}$  &  &  \\ 	   \cline{2-14}

& $^{227}Ac$ \arrow $^{227}Th$                                                                  &     &       & 7.320  & -2.3015 & 6.991 & 6.97  & $7.63\times{10^{-8}}$ & $8.3\times{10^{-8}}$ &  $3.18\times{10^{-11}}$  & $2.9\times{10^{-11}}$ & $2.40\times{10^{3}}$ &  $2.16\times{10^{2}}$  &  &   \\  

& $\left[0.0  , \dfrac{3}{2}^{-}\right]$    $\left[9.3 ,\left(\dfrac{5}{2}^{+}\right)\right]$         & nu & 35.500 & 5.740 & -2.2509 & 7.042 & $\sim$ 7.0 & $7.69\times{10^{-8}}$  &  & $3.21\times{10^{-11}}$  &  & $2.40\times{10^{3}}$ & $2.18\times{10^{2}}$   &  & 28.70 d \\ 
	   
                                                                                     &       &    &      & 6.696 &  -2.2777 & 7.015 &     & $7.56\times{10^{-8}}$  &  & $3.21\times{10^{-11}}$  &  & $2.42\times{10^{3}}$ &   $2.14\times{10^{2}}$ & 21.93 y  &  28.45 d \\ 	   \cline{2-14}

FIG. \ref{Figure8:c} & $^{227}Ac$ \arrow $^{227}Th$                                                                  &     &       & 7.320 & -3.0163 & 6.820 & 6.75  & $8.90\times{10^{-8}}$ & $1.1\times{10^{-7}}$ &  $1.74\times{10^{-15}}$  &  0 & $5.11\times{10^{7}}$ & $8.82\times{10^{2}}$   &  & 28.95 d \\  

& $\left[0.0  , \dfrac{3}{2}^{-}\right]$    $\left[24.5 ,\left(\dfrac{3}{2}^{+}\right)\right]$         & nu & 20.300 & 5.740  & -2.9658 & 6.871 & $\sim$ 6.8 & $8.97\times{10^{-8}}$  &  & $1.74\times{10^{-15}}$  & & $5.15\times{10^{7}}$ & $8.89\times{10^{2}}$   &  &  \\ 
	   
                                                                                       &     &    &      & 6.696 & -2.9925 & 6.844  &      & $8.82\times{10^{-8}}$  &  & $1.74\times{10^{-15}}$  &  & $5.06\times{10^{7}}$ &  $8.74\times{10^{2}}$  &  &  \\ 	   \cline{2-14}

& $^{227}Ac$ \arrow $^{227}Th$                                                                  &     &       & 7.320  & -4.3881 & 6.972 &  - & $4.97\times{10^{-8}}$ & - &  0  & - & $\infty$ &  $1.65\times{10^{4}}$  &  &   \\  

& $\left[0.0  , \dfrac{3}{2}^{-}\right]$    $\left[37.9 ,\left(\dfrac{3}{2}^{-}\right)\right]$         & a & 6.900 & 5.740 & -4.3376 & 7.022  & 6.9 & $5.02\times{10^{-8}}$  &  & 0  &  & $\infty$ & $1.66\times{10^{4}}$   &  &  \\ 
	   
                                                                                      &      &    &      & 6.696  & -4.3643 & 6.995   &     & $4.93\times{10^{-8}}$  &  & 0  &  & $\infty$ &  $1.63\times{10^{4}}$  &  &  \\ 	   \hline\hline

\end{tabular}}

\rule{0pt} {1 cm}
* Experimentally available values are given in the section \ref{3B} of the main text.

\end{sidewaystable*}
\pagebreak[4]
\section{Choice of Spin-Parity for unconfirmed states of neutral atom}\label{spin}

Sometimes the comparison of the calculated $Log~ ft$ values with experimental data gives an idea about the spin-parity of  participating energy levels where these quantities are still  unconfirmed experimentally. 
We have identified a few such transitions in TABLE {\red A.I}. In the transition from $^{123}$Sn $[ 0.0, 11/2^-]$, there are a few states of the daughter $^{123}$Sb, where the spin values are not experimentally confirmed yet (identified as $(J)^\pi$ and/ or $(J^\pi)$ in the table). In the transition from  $^{123}$Sn $[ 0.0, 11/2^-]$ to $E= 1181.3$ keV state of the daughter, if it chooses the decay channel with the spin-parity $J^\pi = (9/2)^+$ then the transition will be of the type (nu), whereas for the choice of spin  $J^\pi = (7/2)^+$, the transition $[0.0, 11/2^- \rightarrow 1181.3, (7/2)^+]$  will be the (u) type. Now comparing with the available experimental $Log ~ft$ value, it seems  from our calculation that the (nu) case is in  good agreement  whereas the (u) case deviates (difference $ \sim 0.4 $ ) from the same for all choices of the radius R.
 
Similarly, from  TABLE  {\red A.I}, our observations for other such transitions are given by (see the table for $Log ~ft$ comparison)\\
 $\bullet$ $^{123}$Sn $[ 0.0, 11/2^-]  \rightarrow ^{123}$Sb $[1260.9, (9/2)^+]$: (nu), \\
 $\bullet$ $^{123}$Sn $[ 0.0, 11/2^-]  \rightarrow ^{123}$Sb $[1337.4, 9/2^+]$: (nu),\\
 $\bullet$ $^{152}$Eu $[ 45.5998, 0^-]  \rightarrow ^{152}$Gd $[1460.5, 1^+]$: (nu).

 Note-1: This type of study is not conclusive in the transition from  $[137.9, 5^-, 6^-]$ level of the $^{148}$Pm nucleus. Depending on the spin of the parent level $5^-/6^-$, all four transitions to the daughter level will either be of type (a) or (nu) and thus, the $Log~ ft$ value in each case will remain the same. 

Note-2: In case of $^{152}$Eu $[45.5998, 0^-]$ to $^{152}$Gd $[1047.9, 0^+]$ transition our $Log ~ft$ differs from that of the experimental value \cite{nndc} by a difference of $\sim 0.16$ - $0.18$ (for different radii). However, we find a numerical mismatch in the  tabulation for the experimental energy value of $[1047.9, 0^+]$ state of $^{152}$Gd \cite{nndc,152}.   

 Note-3: For the $\beta^{-}$ transitions from the first excited state of $^{95}$Nb, the parent level is mentioned as $[234.7, 1/2^{-}]$ in Ref. \cite{NDS95}, whereas in Ref. \cite{nndc} this energy level is mentioned both as $[235.7, 1/2^{-}]$ and $[234.7, 1/2^{-}]$ at different places. However, our calculation of $Log ft$ matches with the reported $Log ft$ value only when we have taken the level energy as 234.7 keV.

\pagebreak[4]

\end{document}